\documentclass{article}

\usepackage{newunicodechar}

\newunicodechar{≤}{\ensuremath{\leq}}

\usepackage{color}   
\usepackage{hyperref}
\hypersetup{
    colorlinks=false, 
    linktoc=all,     
    linkcolor=blue,  
}
\usepackage{caption}
\usepackage{subcaption}

\usepackage[ruled,vlined]{algorithm2e}
\usepackage[english]{babel}
\usepackage{amsmath}
\usepackage{verbatim}

\usepackage[autostyle]{csquotes}
\usepackage{amsfonts}
\usepackage{amssymb}
\usepackage{float}
\makeatletter
 \usepackage{url}

\makeatother

\usepackage{geometry}

\usepackage{graphicx}
\usepackage{multicol}
\usepackage[version=4]{mhchem}
\usepackage[table,xcdraw]{xcolor}
\usepackage{float}

\title{Influence and Influenceability: Global Directionality in Directed Complex Networks}
\author{Niall Rodgers, Peter Ti\v{n}o and Samuel Johnson}
\date{\today}

\begin{document}

\maketitle


\begin{abstract}

Knowing which nodes are influential in a complex network and whether the network can be influenced by a small subset of nodes is a key part of network analysis.  However\textcolor{black}{,} \textcolor{black}{many} traditional measures of importance focus on node level information without considering the global network architecture. We use the method of Trophic Analysis to study directed networks and show that both `influence' and `influenceability' in directed networks depend on the hierarchical structure and the global directionality, \textcolor{black}{as measured by the trophic levels and trophic coherence, respectively}.  We show that in directed networks \textcolor{black}{trophic} hierarchy can explain: the nodes that can reach the most others; where the eigenvector centrality localises; which nodes \textcolor{black}{shape} the \textcolor{black}{behaviour} in \textcolor{black}{opinion or oscillator} dynamics; and which \textcolor{black}{strategies will be successful in} generalised rock-paper-scissors games.  We show, \textcolor{black}{moreover,} that these phenomena are mediated by the global directionality. 
\textcolor{black}{We also highlight} other structural properties of real networks related to influenceability, \textcolor{black}{such as} the pseudospectra, which \textcolor{black}{depend on} trophic coherence. These results apply to any directed network and the principles highlighted -- that node  hierarchy is \textcolor{black}{essential} for \textcolor{black}{understanding} network influence, mediated by global directionality -- \textcolor{black}{are} applicable to many real-world dynamics.

\end{abstract}

\section{Introduction}

Influence in directed complex networks and the ability of the networks to be influenced is vitally important in many real-world systems, for example the spreading of opinions and epidemics or the synchronisation of the brain in a seizure \cite{Lopes2020TheSeizures}. However, when we think of what makes nodes influential, the answer is very disparate and depends on many factors. This question has been of great interest to the network \textcolor{black}{science} community and well studied in \textcolor{black}{the case of} undirected networks \cite{Morone2015InfluencePercolation,Kitsak2010IdentificationNetworks}. However, in the directed case insight can be gained by considering properties unique to \textcolor{black}{directed networks}. This is necessary as many \textcolor{black}{real-world} systems are intrinsically directed and the influence of nodes is strongly tied to the directionality of the \textcolor{black}{edges} \cite{Beguerisse-Diaz2014InterestRiots}. Influence can be thought of as how well nodes control the dynamics of a network, how measures of centrality \textcolor{black}{are} distributed across the network, how many nodes can be reached from a set of nodes and how sensitive the network is to perturbations. We show that all of these properties can be understood and made intuitive by considering the hierarchical ordering and global directionality of the network as measured through the technique of Trophic Analysis \cite{MacKay2020HowNetwork,Rodgers2022NetworkNetworks,Rodgers2023StrongNetworks}.

When the nodes are easily ordered in a hierarchical fashion, \textcolor{black}{as in a food-web,} it is clear that the network can be influenced by the nodes at the bottom of this hierarchy and conversely it is very difficult to influence the network from the top of the hierarchy. When there is little hierarchical structure, \textcolor{black}{ like in an Erd\H{o}s-R\`{e}nyi random graph,} this effect is damped and influence over the network is more evenly distributed. This simple construction, \textcolor{black}{detailed in the background,} aides in the interpretation of control over network dynamics, spreading processes, localisation of centrality measures and sensitivity to perturbations. \textcolor{black}{Hierarchical ordering can explain why nodes which locally look unimportant may be able to influence the entire network.} This framework may provide an interpretation to observations made about influence in directed networks in a variety of literature settings such as the effect of heterogeneous centrality and directionality in opinion formation in real networks \cite{Eom2015OpinionNetworks}, the influence of peripheral nodes on dynamics of directed networks \cite{Wright2019TheDynamics} or the asymmetry between paths up and down the network hierarchy \cite{DelaCruzCabrera2019AnalysisExponential}. Trophic Analysis pairs a global measure of directionality with a local measure of hierarchy, \textcolor{black}{making it different from a single centrality measure.} This allows an intuition surrounding the variability in the importance of a node's position \textcolor{black}{in the hierarchy and how this affects} its role in the network. \textcolor{black}{This} differs from previous results on the relationship between hierarchy and influence which do not feature this pairing of local and global measures \cite{Masuda2009AnalysisNetworks,Masuda2010Dynamics-basedNetworks}. The fact that centrality measures play different roles dependent on the mesoscale structure of the network has been noted previously in the case of PageRank and clustering \cite{Chen2013IdentifyingClustering}.

This paper is organised as follows. We first introduce the background and explain the principles underlying Trophic Analysis. We then introduce some dynamics and highlight how global directionality can affect the influence and  influenceability \textcolor{black}{of these processes}. These are Majority Vote, Kuramoto oscillators, the Voter Model and the frequency of strategies in  Generalised Rock-Paper-Scissors games. We then include some results on the \textcolor{black}{relationship between }structure and \textcolor{black}{influence in}  \textcolor{black}{real-world} networks and how this can be shaped by hierarchy. We demonstrate the relationship between hierarchy and eigenvector localisation, left and right eigenvector correlation  (with \textcolor{black}{specific} real-world examples), sensitivity to structural perturbation through the pseudospectra and the size of the out-component of a node. The real network data used in this paper, \textcolor{black}{also used in \cite{Johnson2017LooplessnessCoherence,Rodgers2023StrongNetworks},} is available at \cite{DataSamJohnson} and contains food-webs, neural networks, social networks and more as well as the original sources of the network data used. \textcolor{black}{This shows how many diverse notions of influence can be investigated in directed networks using Trophic Analysis and how in a wide range of systems the ability to exploit hierarchy to influence a system depends on the global directional organisation.}

\section{Background}

Real-world systems formed by many interacting elements can be represented using graphs. These complex networks are sets of nodes or vertices representing the  elements of the system while the edges or links represent the interactions or connections between elements. Many real-world systems such as social networks, food-webs, the internet and more have interactions which are intrinsically directional and \textcolor{black}{may represent the influence a node has over another \cite{Johnson2020DigraphsSystems}}. This structure can be represented through a matrix, $A$, known as an adjacency matrix  where the edges are represented by the non-zero entries of the matrix. For an \textcolor{black}{unweighted directed} network of $N$ nodes this $N \times N $ matrix is defined as  \begin{equation} \label{Adj_eq}
A_{ij}=
    \begin{cases}
     1  \text{ \quad if there exists an edge } i \to j \\ 
     0  \text{ \quad  otherwise }
    \end{cases}.
\end{equation}
In the case of directed networks this matrix is not symmetric, unlike the undirected case where edges go in both directions and the matrix is symmetric.
\textcolor{black}{Each node $i$ then has an in-degree, $k_i^{in}=\sum_j A_{ji}$, and an out-degree, $k_i^{out}=\sum_j A_{ij}$}.
\textcolor{black}{The matrix $A$} can also be weighted to show the strength of interactions but here we focus on the simple unweighted case although it is possible to use Trophic Analysis in the weighted regime \cite{MacKay2020HowNetwork}.

\subsection{Trophic Analysis}

Trophic Analysis is a technique to quantify the global directionality inherent in real directed networks \cite{MacKay2020HowNetwork}, and is applicable to any directed network. Trophic Analysis was originally derived from ecology \cite{Johnson2014TrophicStability} \textcolor{black}{where the original definition} linked hierarchy to weighted steps from the basal nodes (vertices of in-degree zero), which is an intuitive way to view hierarchy but \textcolor{black}{cannot be} generalised to any directed network \textcolor{black}{without basal nodes} like the definition used here and in \cite{MacKay2020HowNetwork,Rodgers2022NetworkNetworks,Shuaib2022TrophicInformation,Rodgers2023StrongNetworks}. Much of the previous work which applied trophic level and incoherence used the previous definition \cite{Johnson2017LooplessnessCoherence,Johnson2014TrophicStability,Pilgrim2020OrganisationalAnarchy,Klaise2017TheWebs,Klaise2016FromProcesses}. \textcolor{black}{This was successfully applied in} a wide variety of settings including infrastructure \cite{Pagani2019ResilienceNetworks,Pagani2020QuantifyingNetworks}, the structure of food webs \cite{Klaise2017TheWebs}, spreading processes such as epidemics or neurons firing \cite{Klaise2016FromProcesses}, and organisational structuring \cite{Pilgrim2020OrganisationalAnarchy}.

Trophic Analysis combines two parts: the node level \textcolor{black}{local} information, Trophic Level, and a measure of the global network directionality, Trophic Incoherence. Trophic level is a node level quantity which measures where a specific node sits in the network hierarchy. Trophic level is calculated by solving the $N \times N $ matrix equation\textcolor{black}{, first proposed in \cite{MacKay2020HowNetwork},} given by \begin{equation} 
    \Lambda h = v,
    \label{eq_h}
\end{equation}
where $h$ is the vector of trophic levels for each node and $v$ is the vector imbalance of the in and out degree of each node, where each element is defined as $ v_i = k_{i}^{in} - k_{i}^{out}$.
$\Lambda$ is the \textcolor{black}{}Laplacian matrix:
\begin{equation}
    \Lambda = diag(u) - A - A^{T},
    \label{eq_Lambda}
\end{equation}
\textcolor{black}{where $u$ is the sum of the in and out degrees of each node, $ u_i = k^{in}_i + k^{out}_i$, and $A^{T}$ is the transpose of the adjacency matrix, $A$.} The solution of equation \ref{eq_h}, \textcolor{black}{which provides the trophic levels of the nodes}, is only defined up to a constant vector so we take the convention that the lowest level node is set to trophic level zero. The Laplacian matrix \textcolor{black}{is singular by definition, yet a solution can be found either by choosing a node (say $i=1$) and setting its value (e.g. $h_1=0$); or iteratively, which is convenient for very large networks \cite{MacKay2020HowNetwork}.}

In a balanced network, \textcolor{black}{for instance} a directed cycle, there is no hierarchical structure so every node has the same level. \textcolor{black}{This is due to the dependence of equation \ref{eq_h} on the imbalance vector, $ v_i = k_{i}^{in} - k_{i}^{out}$, which means that when the in and out-degrees of all nodes are equal the right side of the equation goes to zero.} In a network with a perfect hierarchy like a directed line, the nodes are assigned integer levels with steps of one between connected nodes. The level distributions of real networks are more complex and lie somewhere between \textcolor{black}{these} extreme cases.

Trophic Incoherence is a global parameter which measures how \textcolor{black}{well} hierarchically ordered the network is. \textcolor{black}{It is} related to the amount of feedback in the system, \textcolor{black}{and thus to} network properties such as the spectral radius, non-normality and strong connectivity \cite{MacKay2020HowNetwork,Rodgers2023StrongNetworks}.
\textcolor{black}{It} is quantified via the trophic incoherence  parameter $F$, which is defined as \begin{equation}
    F = \frac{\sum_{ij}A_{ij}(h_{j} -h_{i}-1)^2}{\sum_{ij}A_{ij}}.
    \label{eq_F}
\end{equation} 

\textcolor{black}{In \cite{MacKay2020HowNetwork}, the authors begin with equation (\ref{eq_F}) (which is the original definition of trophic incoherence \cite{Johnson2014TrophicStability}) and define the trophic levels, $h$, as those which minimise $F$. Hence, the linear form of equation (\ref{eq_Lambda}) and dependence on the imbalance vector comes from the minimisation of the quadratic function in equation \ref{eq_F}.}

Trophic Incoherence measures how far the mean square of the difference in trophic level\textcolor{black}{, $h$ calculated via equation \ref{eq_h},} between start and end vertices of all the edges differs from one.
The equation for Trophic Level, \textcolor{black}{equation \ref{eq_h}}, can be derived by minimising equation \ref{eq_F} with respect to $h$. 
\textcolor{black}{The Trophic Incoherence} takes values between 0 and 1 with real networks found on a spectrum between these extreme values. Networks with a perfect hierarchy are coherent and have $F=0$. When the network is balanced like a directed cycle then $F=1$. It is also possible to speak in terms of coherence instead of incoherence by using the quantity $1-F$. \textcolor{black}{When we talk about hierarchy we refer to the bottom of the hierarchy as nodes of low trophic level and the top of the hierarchy as the nodes of high trophic level. In a directed path the low trophic level nodes would be the start of the path and the high level nodes would be near the end of the path. In a food-web the bottom nodes are plants and the top nodes are apex predators. This, however, is just a convention and the notion of top and bottom can be flipped by taking the transpose of the adjacency matrix \textcolor{black}{(for example, when describing hierarchies of information flow from nodes to nodes)}.}  

\textcolor{black}{Using directionality and hierarchy to analyse the structure of directed networks in this way is quite natural and as such alternative similar formulations exists which rank nodes and measure the global directionality. SpingRank \cite{DeBacco2018ANetworks} views the ranking problem as minimising the energy of a network of directed springs, leading to a similar minimisation problem as the one used in Trophic Incoherence. However, these authors focus on the node level ranks rather than the global directionality and add a regularisation term to remove the invariance of their rank equation to addition of a constant vector. There is also a methodology based on Helmholtz–Hodge decomposition for measuring ``circularity'' in directed networks \cite{Kichikawa2019CommunityNetwork,Iyetomi2017InternationalView}  which has been shown to be equivalent to Trophic Analysis \cite{MacKay2020HowNetwork}.}

\subsection{Network Generation} \label{Net_gen}

In order to \textcolor{black}{study how} the properties of networks depend on their trophic structure it is necessary to be able to numerically generate networks which span the full range of trophic incoherence, \textcolor{black}{while keeping the number of nodes and edges fixed}. We adapt the Generalised Preferential Preying Model (GPPM)  \cite{Klaise2016FromProcesses} in an identical way to \cite{Rodgers2022NetworkNetworks} to the definition of trophic level which does not require basal nodes \textcolor{black}{(vertices of zero in-degree)} and use that to generate the networks we require. This works by generating an initial configuration of $N$ nodes and a small number of \textcolor{black}{random} edges, calculating the initial trophic levels of this setup, adding edges up to the required amount using a probability determined by the trophic level and a temperature-like control parameter which can affect the spread of level differences. In order to generate a network with no basal nodes the initial configuration is that each node has in-degree one with the source vertex for that edge chosen uniformly at random from the other nodes in the network. Once the initial \textcolor{black}{trophic} levels $\Tilde{h}$ are calculated \textcolor{black}{via equation \ref{eq_h}} then new edges, \textcolor{black}{up to the required amount,} can be added with probability defined as  \begin{equation} \label{gen:probablity}
        P_{ij}= \exp{\left[ -  \frac{(\tilde{h}_{j} - \tilde{h}_{i} -1 )^2}{2T_{\text{Gen}}}\right]},  
    \end{equation}     
where $P_{ij}$ is the probability of connecting nodes i to j. The parameter $T_{\text{Gen}}$ controls this process. When this parameter is small it is likely that edges connect only between nodes where the level difference is 1 or near to it. When $T_{\text{Gen}}$ is very large then the \textcolor{black}{edge-addition} probability goes towards 1 irrespective of the trophic level difference \textcolor{black}{between the end and start nodes}.

\begin{figure}[H]
      		\centering
            \includegraphics[width=0.8\linewidth]{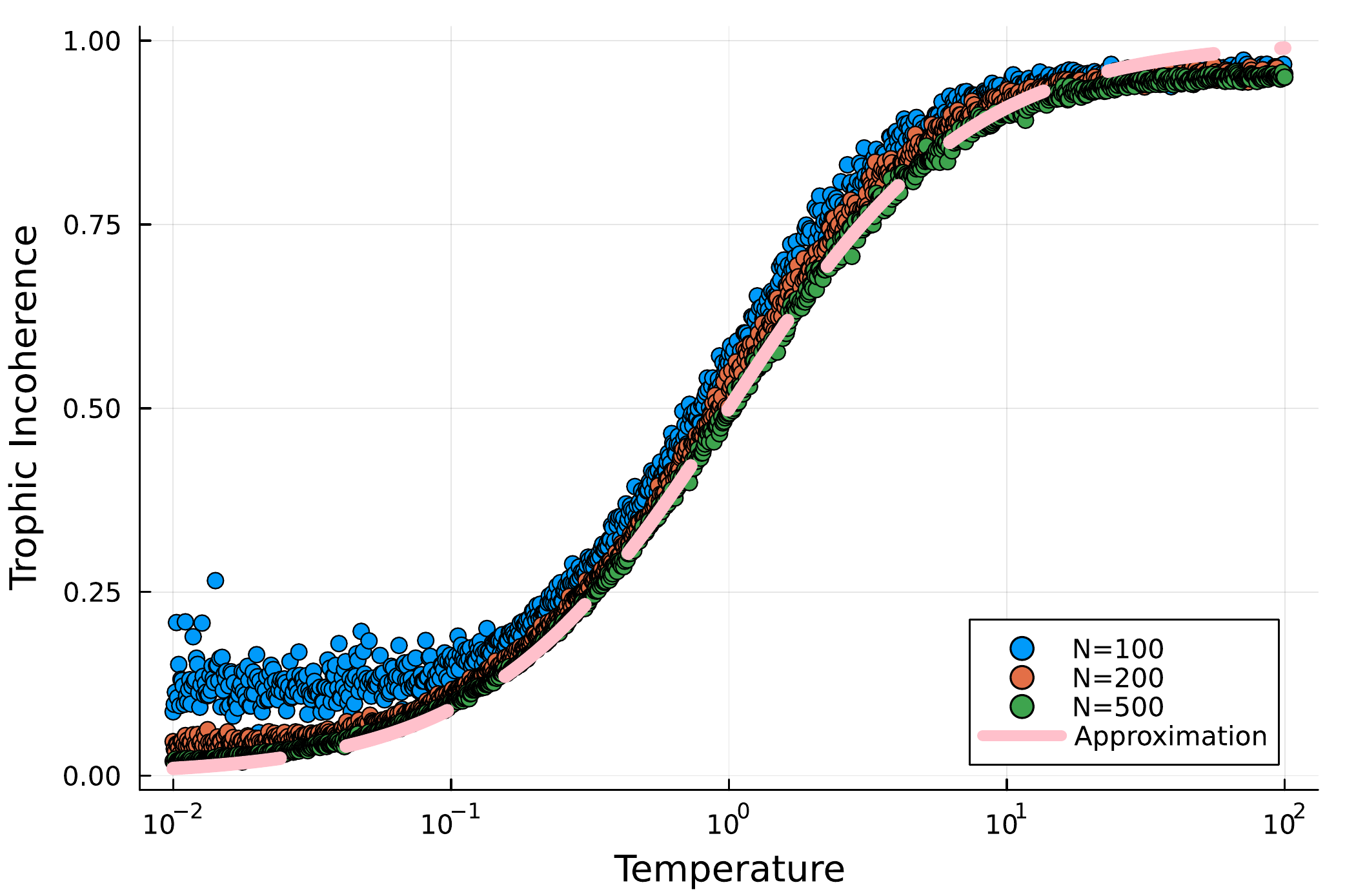}
        	\caption{ \textcolor{black}{Relationship between Trophic Incoherence and generation temperature in the model used in this work, where each of the 1000 points per network size represents a distinct} \textcolor{black}{network generated by the model,} \textcolor{black}{with $N=100$, $N=200$ and $N=500$, $\langle k \rangle = 20$ and no basal nodes. The generation temperature is logarithmically spaced between $10^{-2}$ and $10^2$. We also plot with a dashed line an analytical approximation, given by equation (\ref{eq_Fapprox}), for the relationship between the Trophic Incoherence and Temperature.}  
        	}
        	
        \label{fig:generation_example}
        \end{figure}

\textcolor{black}{The relationship between the generation temperature and the trophic incoherence is displayed in figure \ref{fig:generation_example}, which shows how the model can be used to create a sample of networks of varying incoherence. In order to efficiently sample the whole spectrum of the trophic incoherence we use logarithmic spacing of $T_{\text{Gen}}$ throughout the paper, \textcolor{black}{except for in the sections on the generalised rock-paper-scissors games} and the out-component analysis. In these two sections we compare networks of low, intermediate and high trophic incoherence, for which we selected the generation temperatures to be \textcolor{black}{$0.02,1$ and $100$, respectively}. As shown in figure \ref{fig:generation_example} this picks out each of the broad regimes of the generative model; the low temperature regime of very coherent networks, the intermediate regime where we see more variability with the temperature, and the high temperature regime of incoherent networks.}

\textcolor{black}{We also show in figure \ref{fig:generation_example} how the behaviour of this model can be analytically approximated. In previous work on trophic analysis, where the ecological definition of trophic levels was used, the trophic incoherence, $q$, was defined as the standard deviation of the trophic level differences spanned by edges \cite{Johnson2014TrophicStability}. Under this definition the mean trophic level is always one, and a network is more incoherent the more heterogeneously trophic differences are distributed around this value. Given a set of trophic levels (by whichever definition), it is possible to  convert \textcolor{black}{between the two measures of incoherence} via the formula defined in \cite{MacKay2020HowNetwork}, \begin{equation}
    F = \frac{\eta^2}{1 + \eta^2},
\end{equation}
where $\eta =\sigma/\mu$,  $\sigma$ is the standard deviation of the level differences and $\mu$ is the mean level difference. Hence, $\eta$ is the appropriate replacement for the incoherence parameter $q$ used in \cite{Johnson2014TrophicStability, Johnson2017LooplessnessCoherence} and subsequent papers, when using the MacKay level definition \cite{MacKay2020HowNetwork}. If we assume that the trophic level difference distribution across the edges in the generated networks is distributed as in the edge addition probability, we can replace $\eta^2$
with $T_{\text{Gen}}$,
if we assume that iteration of the attachment probability given by equation (\ref{gen:probablity}) leads to an approximately Gaussian distribution of trophic differences. The differences in trophic level has been found roughly to follow a Gaussian distribution in real networks. \textcolor{black}{Also, analytical results} which hold for real networks have been based on the assumption that the level difference distribution is Gaussian \cite{Rodgers2023StrongNetworks}.   The assumption that the trophic levels follow the same Gaussian as implied by the edge addition probability leads to the approximation that \begin{equation}
    F \approx \frac{T_{\text{Gen}}}{1 + T_{\text{Gen}}}.
    \label{eq_Fapprox}
\end{equation}
As shown by figure \ref{fig:generation_example} this fits quite well for the larger networks sizes.  We plot the data with a logarithmic scaling on the x-axis which allows the different regimes of low, intermediate and high incoherence to be visualised and leads to the sigmoid shape. As writing $T_{\text{Gen}}=\log_{10} a$ leads to the alternative form \begin{equation}
    F \approx \frac{1}{1 + 10^{-a}}\end{equation} which is equation of a sigmoid curve. \textcolor{black}{This approximation fails} at low temperatures for the small very dense networks, $N=100$, as the generative model struggles to make networks which are very \textcolor{black}{coherent at such a high edge density - it becomes difficult to place edges} in locations where the coherence is maximised. We note that at high $T_{\text{Gen}}$ we obtain $\eta^2< T_{\text{Gen}}$, however the approximation works fairly well thanks to the sigmoid function, which is close to one for all large arguments. Hence, $\eta^2$ and $T_{\text{Gen}}$ are not directly equivalent, \textcolor{black}{but relatable} when used to calculate $F$ in this way. Nevertheless, this approximation provides an illustration of how the model works. Incoherence increases with temperature by augmenting the probability that edges are placed with edge difference further from one. }

\textcolor{black}{Additional} detail of how to efficiently generate these networks \textcolor{black}{by more efficiently sampling the space of edge addition probabilities so that each draw of a random number results in an edge being added} is given in \cite{Rodgers2022NetworkNetworks}. \textcolor{black}{All graph manipulations in this work were carried out using Julia Package Graphs.jl \cite{Graphs2021}.}

\section{Influence and Influenceability of Dynamics}

\textcolor{black}{In this section we present a wide range of dynamics and show how trophic level and incoherence can be used to provide insight into disparate and unrelated processes. The idea is to use simple dynamics to highlight the phenomena being picked out by trophic analysis. In this case we define `Influence' to mean the ability of a targeted perturbation or modification of the nodes to affect the state of the system some time after it is applied. In each of the dynamics we show how the nodes which can be regarded as `influential' are the ones with low trophic level -- i.e. those at the bottom of the hierarchy. We also find that the `influenceability' of the whole network by the low trophic level nodes depends on the network's trophic coherence. The goal of this section is to highlight the broad range of dynamics to which trophic analysis can be applied. In this section we use numerically generated networks only \textcolor{black}{- as opposed to} the structural section where we provide results on numerically generated networks and a data-set of real-world networks \cite{DataSamJohnson}.}
\textcolor{black}{We generate networks numerically with the model described above, which allows us to set the numbers of nodes and edges and to vary the trophic coherence. We can also set the number of basal nodes} (nodes with zero in-degree). \textcolor{black}{
The dynamics of basal nodes are not affected by the states of other nodes, since they have no inputs. However, they are obvious candidates for nodes with a high influence on the whole network, and this has been demonstrated in previous work \cite{Wright2019TheDynamics,Johnson2020DigraphsSystems}. While basal nodes will tend to have low trophic levels, trophic analysis is not needed in order to identify them. We therefore fix the number of basal nodes in our generated networks to be zero, so as to highlight the importance of trophic level even in networks where all nodes can be affected by the rest of the system.
}

\textcolor{black}{Our hypothesis is that the nodes at the lowest trophic levels (regardless of whether they are basal nodes) are the most influential across a broad range of different dynamics. We test this in the following way. We set up initial conditions such that all the nodes begin in one state or frequency, except for a `perturbed fraction' of nodes which start in a different state or frequency. We choose the nodes with lowest trophic level for this perturbation, and observe the subsequent dynamics of the system. Whenever the system evolves to the state or frequency initially given to, say, only 5\% of the nodes, we can conclude that: the network is highly `influenceable', and these are the most `influential' nodes. We go on to find that influenceability depends on trophic coherence, and influence is indeed determined by trophic level, across several dynamics.}

We present the results as scatter plots of individual networks, \textcolor{black}{but also show the averages} and standard deviations of the sets of networks we generate. This allows the variability of the dynamics for a small number of influential nodes to be visualised as well as showing the general trend. \textcolor{black}{For each of the dynamics we influence 5\%, 20\% and 40\% of the nodes with the lowest trophic level to show how varying the size of this set affects the influence and the spread of the results.}

\subsection{Majority Vote}

Majority vote dynamics is the first dynamics to be studied as it is very simple to define as well as being widely used. This allows the primary focus to be on the network structure. Each agent \textcolor{black}{(node)} in the model is given an `opinion' and then updates their opinion such that they select the opinion held by the majority of their neighbours. This model is very simplistic \textcolor{black}{yet} shares some properties with other common agent-based discrete dynamics widely used in complex networks, such as the SIS epidemic model \cite{Volkening2020ForecastingInfection}, \textcolor{black}{the} Hopfield neural network model \cite{Rodgers2022NetworkNetworks}, the Ising model \cite{Kim2021Majority-voteNetworks} and the Moran process from biology \cite{Moinet2018GeneralizedAttractiveness}. Each of these models \textcolor{black}{has} additional details which separate \textcolor{black}{it} from the basic majority vote, but the dynamics \textcolor{black}{is} governed by similar principles.

Majority Vote dynamics can be defined as a system where there are two opinions, denoted as +1 and -1. An agent holds either of the two options and updates their state using an update rule defined as
\begin{equation}
 S_{i}(t + \Delta t) = \text{sgn}\left(\sum_{j=1}^N A_{ji}S_j(t) \right).\end{equation} 
The updates can be done asynchronously or in parallel and in this work we chose parallel updates. This specific update rule is deterministic but it can be modified in a range of ways to add stochastic behaviour to the system. Additionally, more than two \textcolor{black}{possible} states could be added to model a larger number of opinions. 
\textcolor{black}{
An absorbing state is reached if all nodes have the same opinion. However, it is also possible for the system to remain in a fixed point or in a cycle involving nodes with different opinions. In practice, we allow the system to run for a specified number of updates before recording the proportion of nodes in the $+1$ opinion.
}


\begin{figure}[H]
     \centering
     \begin{subfigure}[t]{0.45\textwidth}
         \centering
         \includegraphics[width=\textwidth]{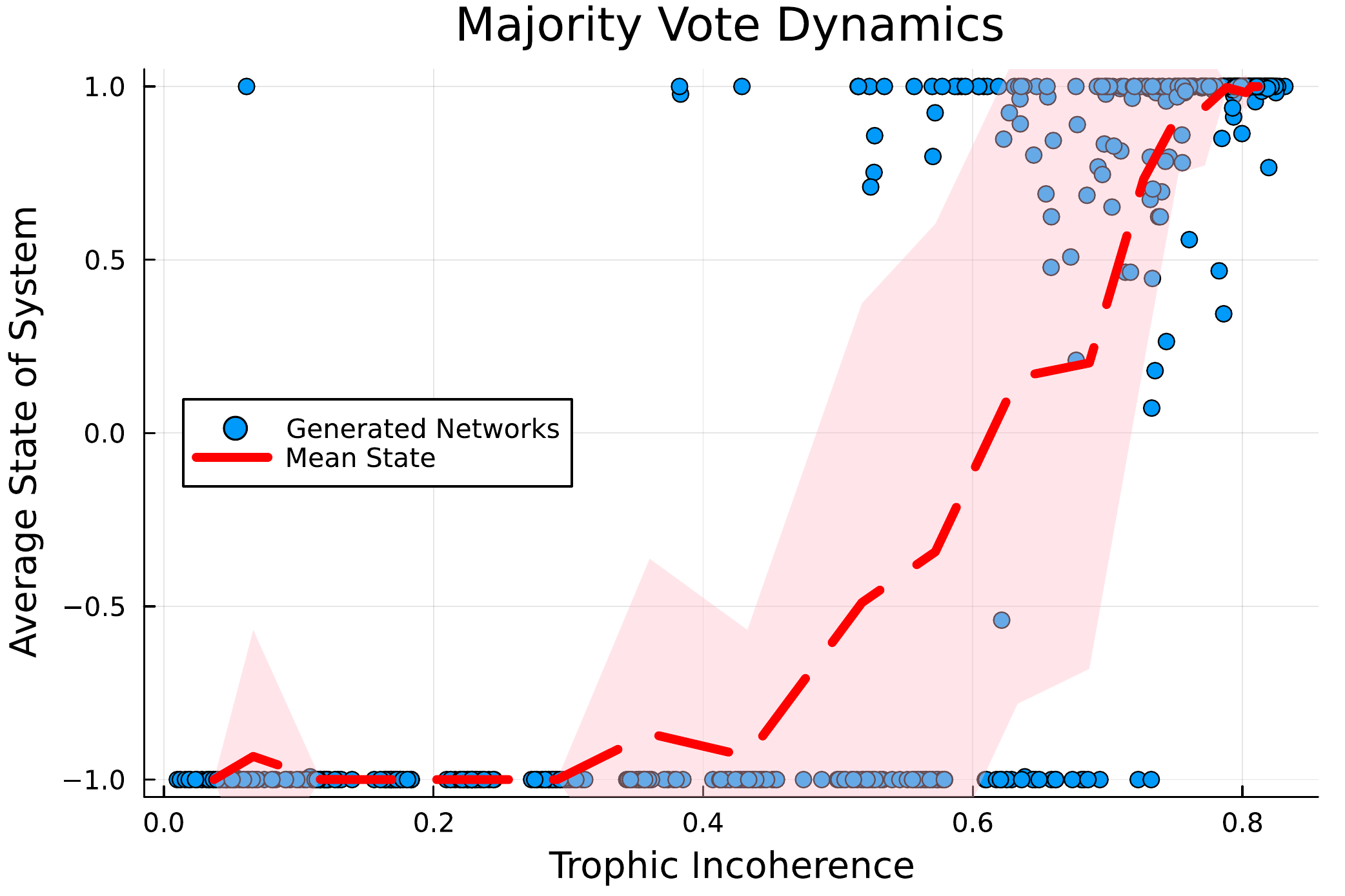}
         \caption{Influence by 5\% of nodes with lowest Trophic Level.}
         \label{fig:maj_5}
     \end{subfigure}
     \hfill
     \begin{subfigure}[t]{0.45\textwidth}
         \centering
         \includegraphics[width=\textwidth]{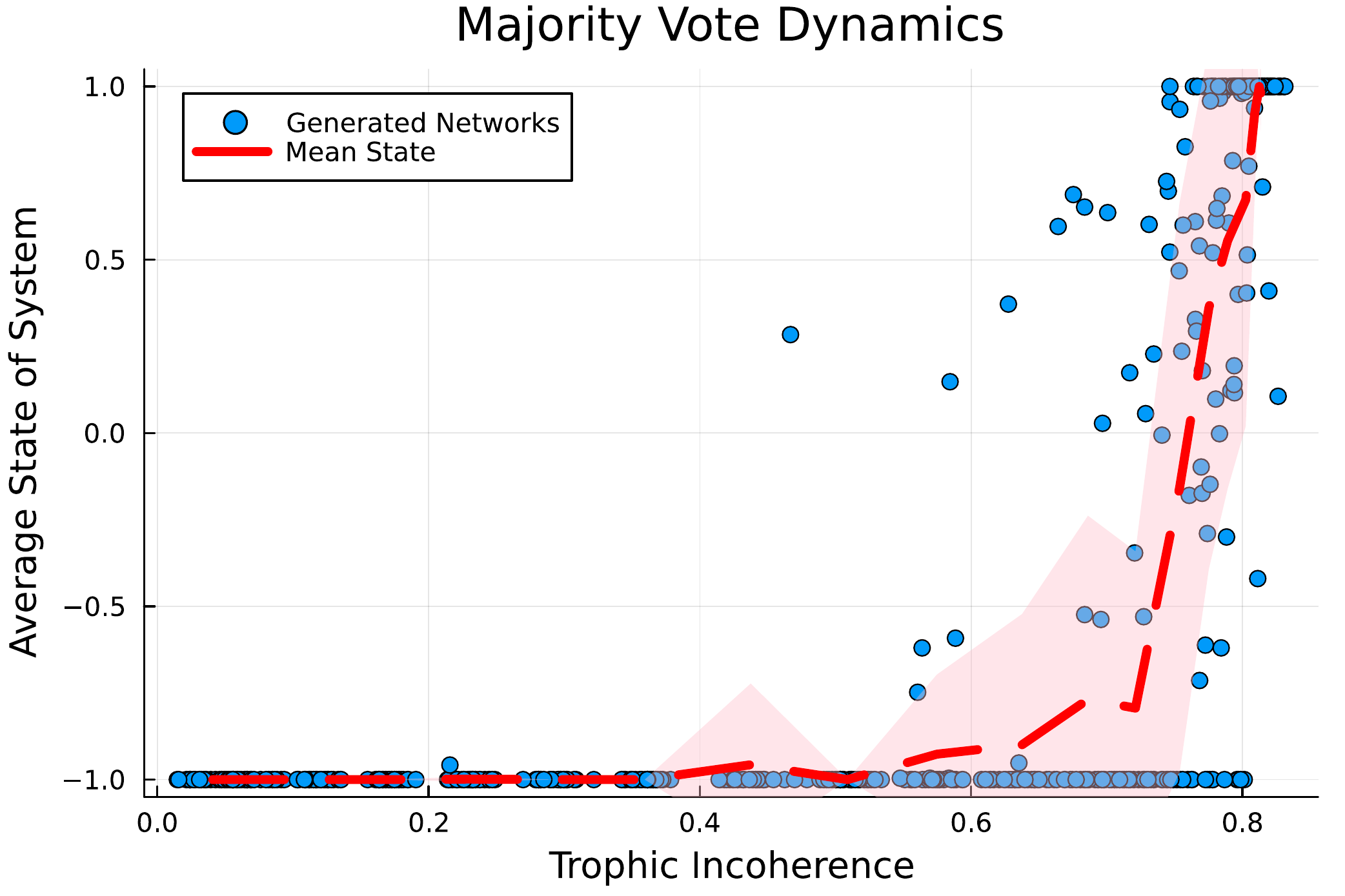}
         \caption{Influence by 20\% of nodes with lowest Trophic Level.}
         \label{fig:maj_20}
     \end{subfigure}
     \hfill
     \begin{subfigure}[t]{0.45\textwidth}
         \centering
         \includegraphics[width=\textwidth]{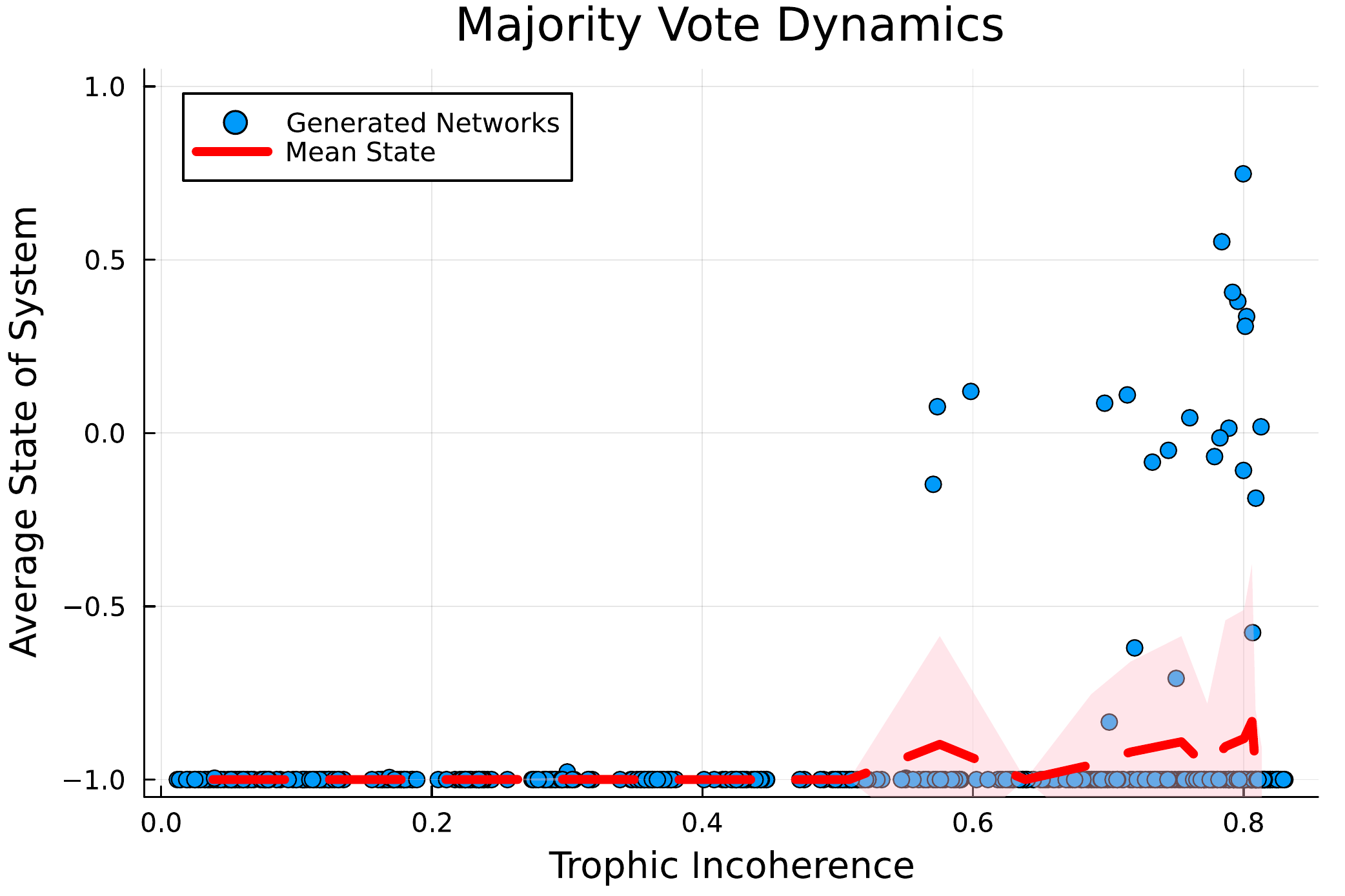}
         \caption{Influence by 40\% of nodes with lowest Trophic Level.}
         \label{fig:maj_40}
     \end{subfigure}
        \caption{Majority Vote Dynamics \textcolor{black}{after 1000 updates when a specified fraction of nodes 
        are initially set to $-1$ with the rest of the nodes set to $+1$. This fraction is 5\%, 20\% and 40\% in panels (a), (b) and (c), respectively. The nodes set to $-1$ are those with lowest trophic level. The average state of the system \textcolor{black}{(the fraction in the +1 opinion)} is shown against trophic incoherence,}\textcolor{black}{with each point a distinct} \textcolor{black}{network generated by the model in section \ref{Net_gen}} \textcolor{black}{with} $N=500$ and $\langle k \rangle = 5$ \textcolor{black}{and} no basal nodes. \textcolor{black}{ Temperatures, $T_{\text{Gen}}$, are logarithmically spaced between $10^{-1}$ and $10^2$ with 20 distinct temperatures used. At each temperature we generate 30 networks and compute the mean and standard deviation of this set with the mean plotted with the red dashed line and the standard deviation shown with the shaded area.}}
        \label{fig:maj_vote_multiple_examples}
\end{figure}

The effectiveness of using trophic level to identify influential nodes is shown in figure \ref{fig:maj_vote_multiple_examples}. When a small number of nodes at the bottom of the hierarchy  (\textcolor{black}{nodes with low trophic level}) are placed in the -1 \textcolor{black}{opinion, most or all the rest of the nodes can evolve towards this state} despite \textcolor{black}{the majority of the nodes being initialised in the +1 opinion}.
\textcolor{black}{This scenario holds except when the networks become too incoherent for the perturbed nodes to have a decisive influence. Viewed as a function of trophic incoherence, there is a transition from an influenceable regime (in which the low trophic level nodes dominate) to one which is less influenceable. Where this transition occurs varies with the fraction of perturbed nodes. Moreover, for a given trophic incoherence and perturbed fraction, some network realisations are highly influenceable and others are not, suggesting that there are other relevant network properties to be determined.}


This is the expected behaviour in networks of lower incoherence as there is less feedback and cycles so the \textcolor{black}{majority} opinion at the bottom of the network
moves up through the hierarchy,
\textcolor{black}{eventually changing the opinion of the whole system.}
In a network which is more incoherent \textcolor{black}{this does not happen, for two reasons: there are fewer nodes connected only to the perturbed fraction; and there is more feedback (from directed cycles), which allows the unperturbed nodes to reinforce their initial opinion.}

 \textcolor{black}{The standard deviation of the points is largest at the incoherence which is between the two regimes. When 40\% of the nodes are perturbed, figure \ref{fig:maj_40}, almost all nodes are influenced by the low level nodes apart from in the most incoherent networks, which may \textcolor{black}{remain} in an intermediate state. Similar behaviour is seen when 20\% of nodes are \textcolor{black}{perturbed}, figure \ref{fig:maj_20}, although there is now a clearer regime of networks which are too incoherent to be influenced by this fraction of nodes. When only 5\% of nodes are \textcolor{black}{perturbed}, fig \ref{fig:maj_5}, despite the number of nodes being small, they are able to \textcolor{black}{influence} the state of many of the networks of low coherence and the average state only increases \textcolor{black}{as the incoherence} becomes intermediate and there are more cycles and feedback in the system. There  are a few outlier points at low incoherence. This is to be expected as we \textcolor{black}{perturb} such a small fraction of the nodes that there may be features of the network which make it difficult for them to influence the system. For example the number of edges which leave the set which is perturbed may be small or the trophic levels may be organised in such a way that the small set does not fill the lowest level of the network \cite{Rodgers2022NetworkNetworks}.} \textcolor{black}{However, 
these simulations still demonstrate the usefulness of this method. Indeed, in very coherent networks specific targeted influence can overcome an opinion held by 95\% of the network.} This example provides \textcolor{black}{perhaps} the simplest demonstration \textcolor{black}{of} the insight trophic analysis can give into network dynamics. This may have potential applications in social networks and organisations, particularly in the promoting of cooperation in prisoner dilemma-like games on networks \cite{Pena2009ConformityDilemma}.

\subsection{Synchronisation Phenomena and Kuramoto Oscillators  }

\begin{figure}[H]
     \centering
     \begin{subfigure}[t]{0.45\textwidth}
         \centering
         \includegraphics[width=\textwidth]{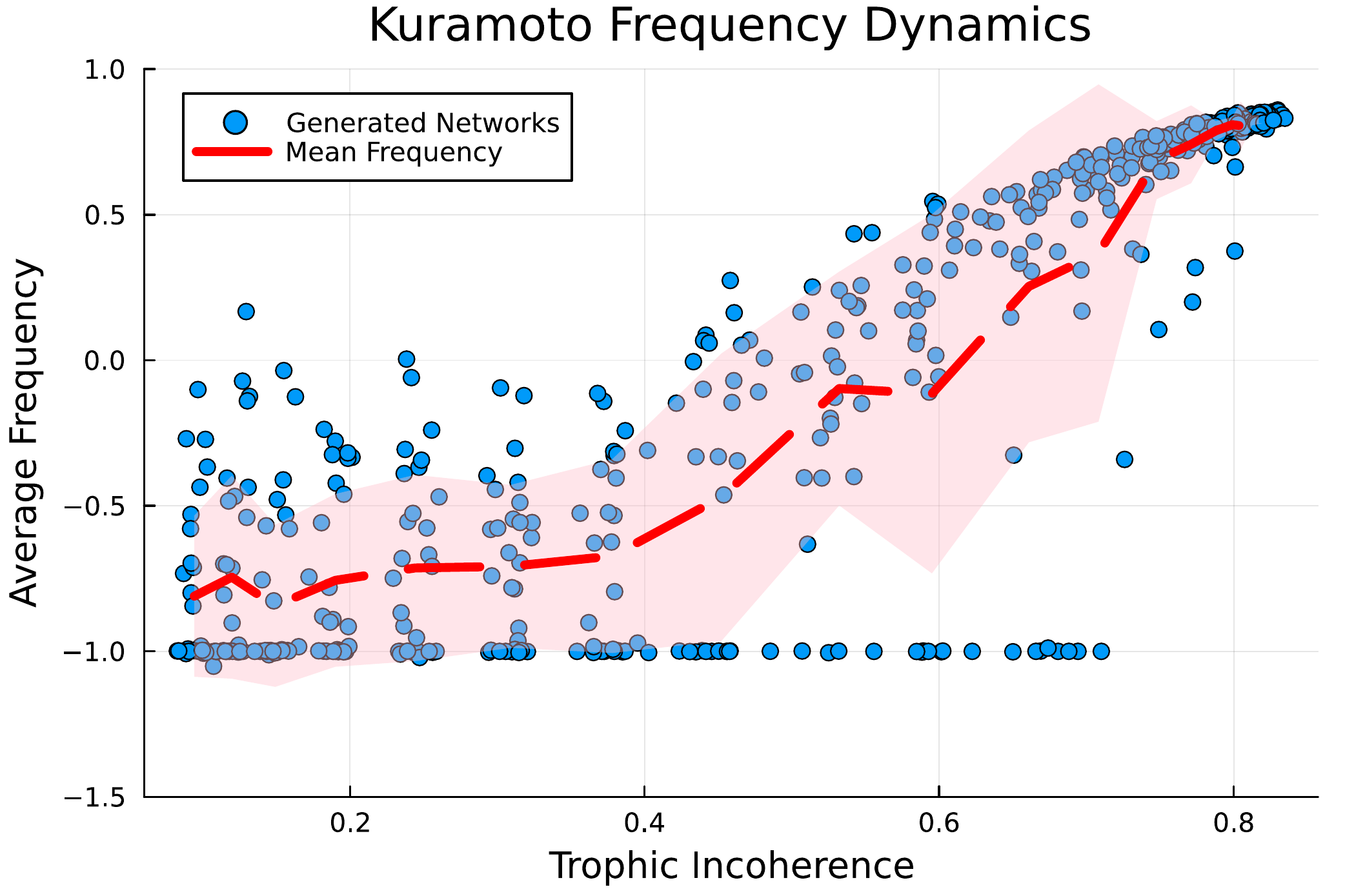}
         \caption{Influence by 5\% of nodes with lowest Trophic Level.}
         \label{fig:kuramoto_5}
     \end{subfigure}
     \hfill
     \begin{subfigure}[t]{0.45\textwidth}
         \centering
         \includegraphics[width=\textwidth]{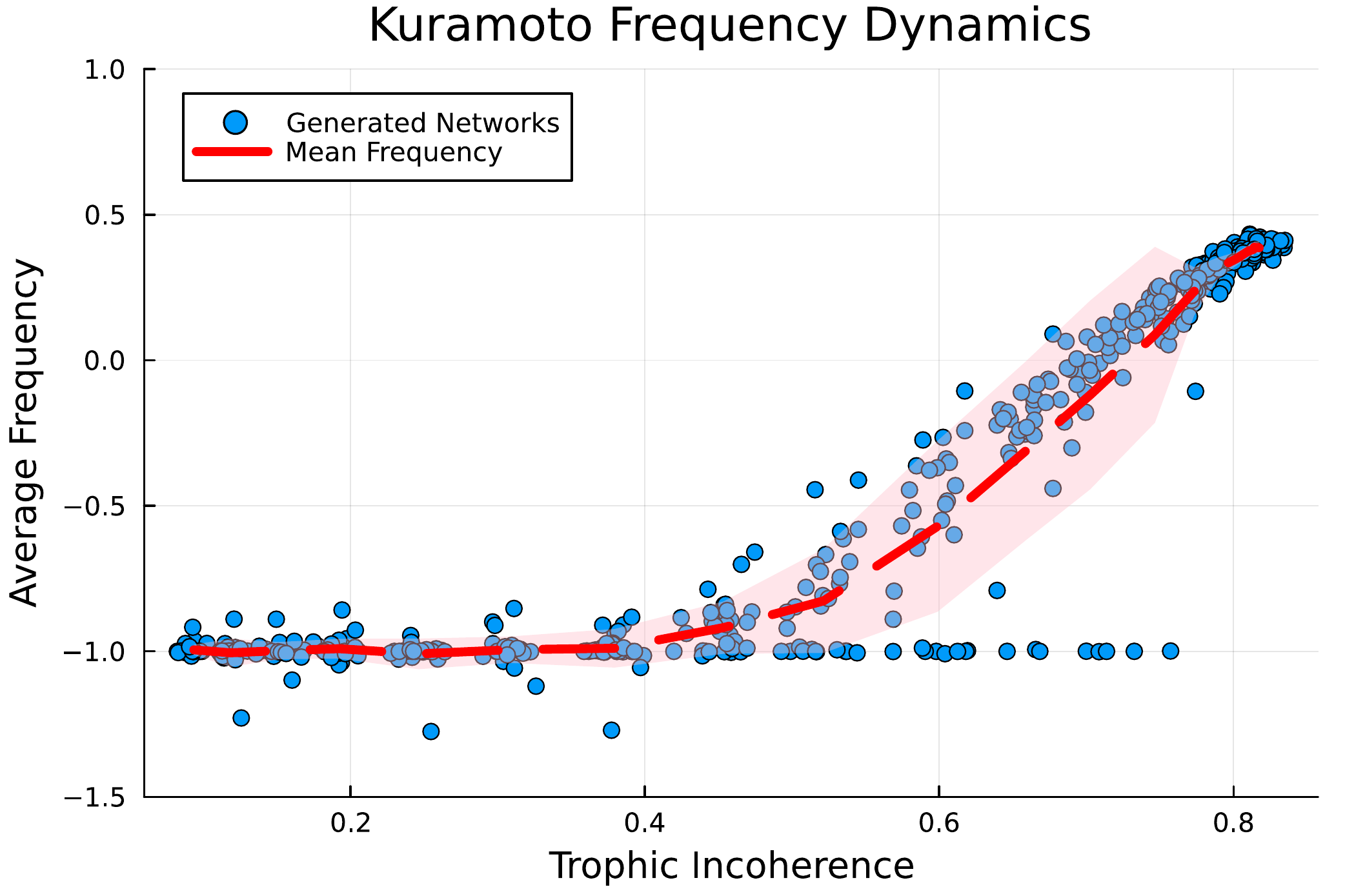}
         \caption{Influence by 20\% of nodes with lowest Trophic Level.}
         \label{fig:kuramoto_20}
     \end{subfigure}
     \hfill
     \begin{subfigure}[t]{0.45\textwidth}
         \centering
         \includegraphics[width=\textwidth]{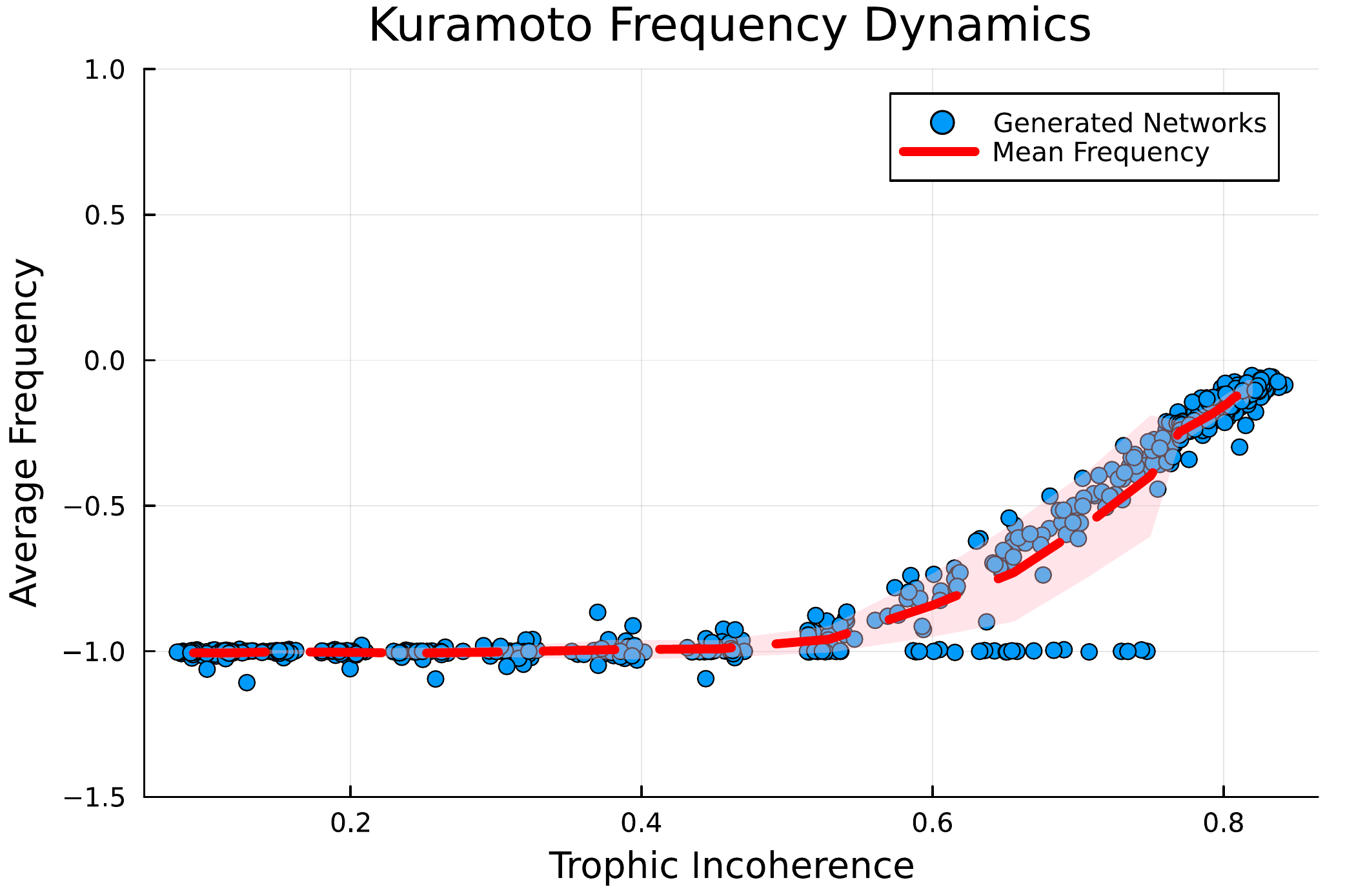}
         \caption{Influence by 40\% of nodes with lowest Trophic Level.}
         \label{fig:kuramoto_40}
     \end{subfigure}
        \caption{Kuramoto Model Dynamics \textcolor{black}{after simulating for 20 time units when different fractions of nodes, chosen by lowest Trophic Level, are \textcolor{black}{given intrinsic frequency} $-1$ while} the rest of the nodes are \textcolor{black}{given intrinsic frequency} $+1$; for varying Trophic Incoherence, \textcolor{black}{where each point is a distinct} \textcolor{black}{network generated by the model in section \ref{Net_gen}} \textcolor{black}{with} $N=500$, $\langle k \rangle = 5$ \textcolor{black}{and} no basal nodes. \textcolor{black}{ Temperatures, $T_{\text{Gen}}$, are logarithmically spaced between $10^{-1}$ and $10^2$ with 20 distinct temperatures used. At each temperature we generate 30 networks and compute the mean and standard deviation of this set, with the mean plotted with the red dashed line and the standard deviation shown with the shaded area. \textcolor{black}{The coupling constant, $K=100$, is twenty times the average degree to ensure synchronisation. Nodes are initially synchronised with phase 0.}} }
        \label{fig:kuramoto_multiple_examples}
\end{figure}

A very similar result to that observed in the discrete majority vote dynamics can be observed in the \textcolor{black}{frequency} synchronisation of continuous Kuramoto oscillators, where the low level nodes can \textcolor{black}{influence} the \textcolor{black}{frequency} of the \textcolor{black}{whole} system. The Kuramoto model is a \textcolor{black}{well-known} model of synchronisation with a wide range of applications, \textcolor{black}{from} neuroscience \textcolor{black}{to power grids} \cite{kuramoto1984chemical,Acebron2005ThePhenomena,Bick2020UnderstandingReview, Cumin2007GeneralisingBrain,guo2021overviews}. We use NetworkDyanmics.jl \cite{Lindner2021NetworkDynamics.jlComposingJulia} to solve the system of differential equations used in our variant of the model. \textcolor{black}{Each oscillator (node) $i$ has a phase, $\theta_i$, }which evolves according to the equation \begin{equation}
    \frac{d\theta_i}{dt } = \omega_i + \frac{K}{k_{i}^{\text{in}}}\sum_{j=1}^N A_{ji}\sin{(\theta_j - \theta_i)},
\end{equation}   
where $K$ is the coupling constant, $k_{i}^{\text{in}}$ is the in-degree of the node \textcolor{black}{and $\omega_i$ is the natural frequency of the node.} We use the form normalised by in-degree so that the oscillators update at similar rates \textcolor{black}{regardless of the number of} input nodes. For the system  used we set the coupling constant to \textcolor{black}{$K = 20 \langle k \rangle $,} which is well above the critical threshold \textcolor{black}{for maintaining frequency and phase synchronisation when the system is initialised at a single phase.} This equation could be modified by adding noise or delays to the updates to make the system more realistic. However, we 
\textcolor{black}{focus here on the simplest case.}
Phase Synchronisation can be measured by the order parameter \begin{equation}
    r = \frac{1}{N} \bigg\lvert \sum_{i=1}^N  e^{i\theta_i} \bigg\rvert,
\end{equation} 
which reaches one when all the oscillators share the same phase but is zero when the phases are uniformly randomly distributed from 0 to $2\pi$. How the low \textcolor{black}{trophic} level nodes \textcolor{black}{influence} the \textcolor{black}{frequency} in networks \textcolor{black}{with varying trophic coherence} is \textcolor{black}{shown} in figure \ref{fig:kuramoto_multiple_examples}, where \textcolor{black}{a fraction of} the low \textcolor{black}{trophic} level nodes \textcolor{black}{have natural frequency $\omega=-1$} \textcolor{black}{and the remainder \textcolor{black}{have $\omega=+1$}}. \textcolor{black}{Average frequency is measured by sampling differences in node phase during the latter half of the simulation and  dividing by the time differences, averaging over samples and nodes. Since we are interested in which frequency the nodes synchronise to, we set the initial condition so that the nodes begin with the same phase and the dynamics are driven by the topology and natural frequencies only, and not the varying initial conditions. The networks maintain phase synchronisation, $r$ close to 1, and adequate frequency synchronisation, standard deviation of node frequencies is generally very small and always less than 1 (equivalent to standard error less than 0.045), as the simulation evolves due to the initial conditions and strong coupling.} Networks with low incoherence propagate \textcolor{black}{the frequency of the low-level nodes} through the whole system and the network synchronises at that {frequency}. In  networks of high trophic incoherence the feedback from the higher level \textcolor{black}{nodes back to the lower} level nodes dilutes this effect, \textcolor{black}{so} the final average {frequency} of the system is further from that at the low level nodes \textcolor{black}{and closer to that of the higher level nodes}.  

This transition is different from the majority vote transition \textcolor{black}{since both phase and frequency are} continuous, rather than the discrete states available to the majority vote dynamics. \textcolor{black}{Additionally, in this case we are permanently changing an intrinsic property of the nodes which is maintained throughout the process, whereas previously the perturbation was to the initial conditions.} However, the fact that applying the same principle to two very different systems \textcolor{black}{results in qualitatively similar behaviour} highlights the usefulness of Trophic Analysis as a tool. \textcolor{black}{When only 5\% of the low level nodes are selected in the Kuramoto case, figure \ref{fig:kuramoto_5}, the trend is noisy and the standard deviation is large. However, even in this regime we see the trend of the more coherent networks being more easily influenced by the low level nodes and only the most incoherent networks resistant to this. When 20\% of the low level nodes are set, figure \ref{fig:kuramoto_20}, the trend is much clearer. The most coherent networks are all very strongly influenced by the low level nodes, and there is a clear transition as the network becomes more incoherent, with a lower standard deviation. When 40\% of the nodes are perturbed, figure \ref{fig:kuramoto_40}, the system is more strongly influenced by the low level nodes \textcolor{black}{and the frequency is reduced away from one, even in the most incoherent networks.}
In all cases, even at high incoherence there are some networks which synchronise with the `influenced' (low level) nodes. 
}

\textcolor{black}{
The oscillators begin in a synchronised state,
so we find similar results when the end time of the simulation is varied. If a different oscillator function were used it may be possible to exploit the network hierarchy to create more complex states. It has recently been shown that oscillators on non-normal networks can lead to chimera states \cite{muolo2023persistence}, and it is known that non-normality is related to trophic coherence \cite{Johnson2020DigraphsSystems,MacKay2020HowNetwork}.
}

\subsection{Voter Model}

A Voter model is another way to model opinion formation for which similar results around influence and influenceability can be demonstrated. Similar to the majority vote dynamics the agents can be given states of either +1 or -1 and updated according to an update rule. However, instead of taking the majority opinion of the neighbours the agent can update its state by choosing at random one of its incoming neighbours and adopting the selected state. This model can be adapted,  increased in complexity and then applied to study real voting processes \cite{Fernandez-Gracia2014IsVoters}, economics \cite{Kirman1993AntsRecruitment} and chemistry \cite{Fichthorn1989Noise-inducedModel,Considine1989CommentModel}. \textcolor{black}{The model has two absorbing states where all nodes share the same opinion (consensus). However, after a finite time the system can find itself in an intermediate state which includes both opinions (and on a directed network the absorbing states can be inaccessible).}

\begin{figure}[H]
     \centering
     \begin{subfigure}[t]{0.45\textwidth}
         \centering
         \includegraphics[width=\textwidth]{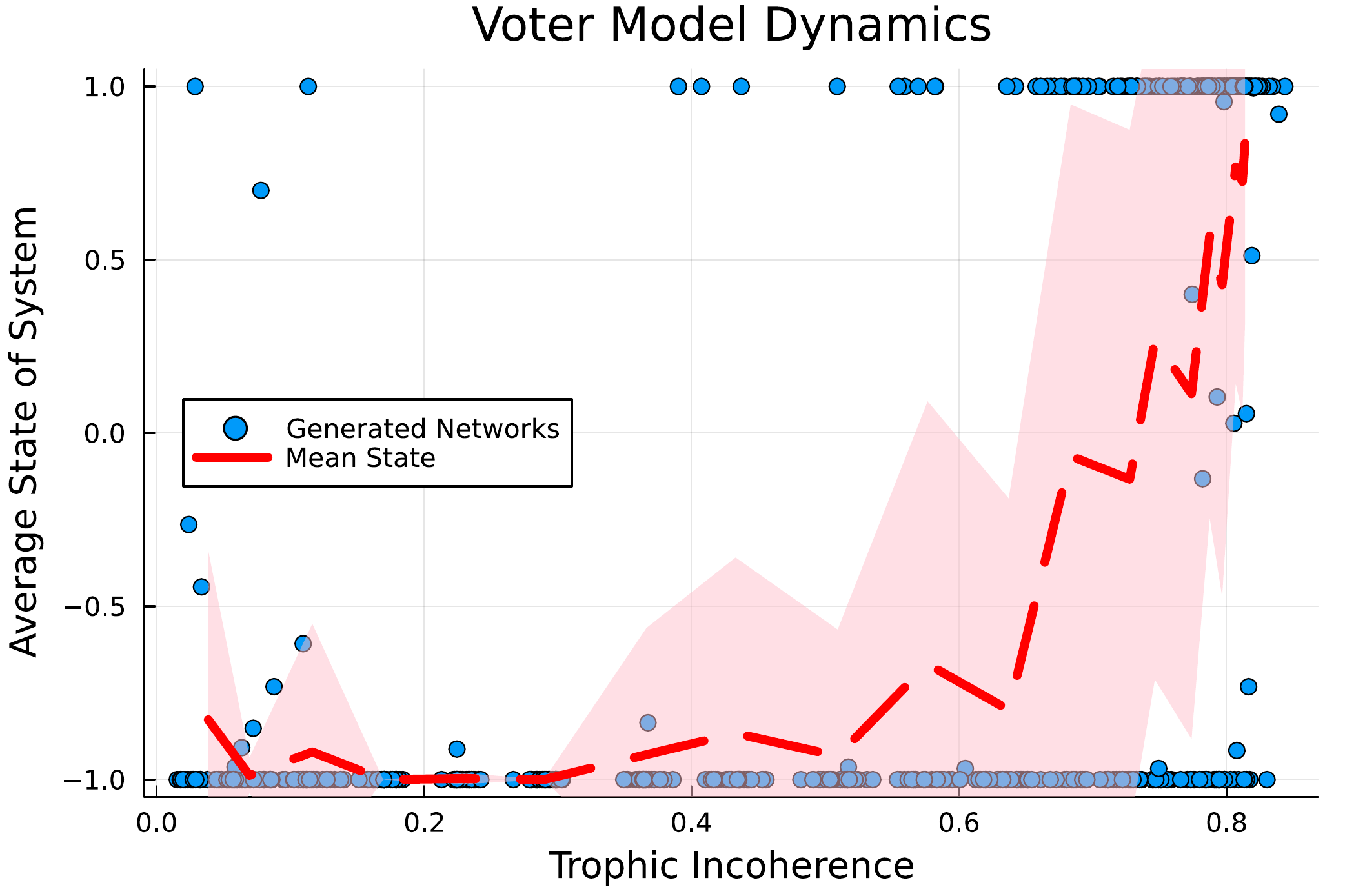}
         \caption{Influence by 5\% of nodes with lowest Trophic Level.}
         \label{fig:vote_5}
     \end{subfigure}
     \hfill
     \begin{subfigure}[t]{0.45\textwidth}
         \centering
         \includegraphics[width=\textwidth]{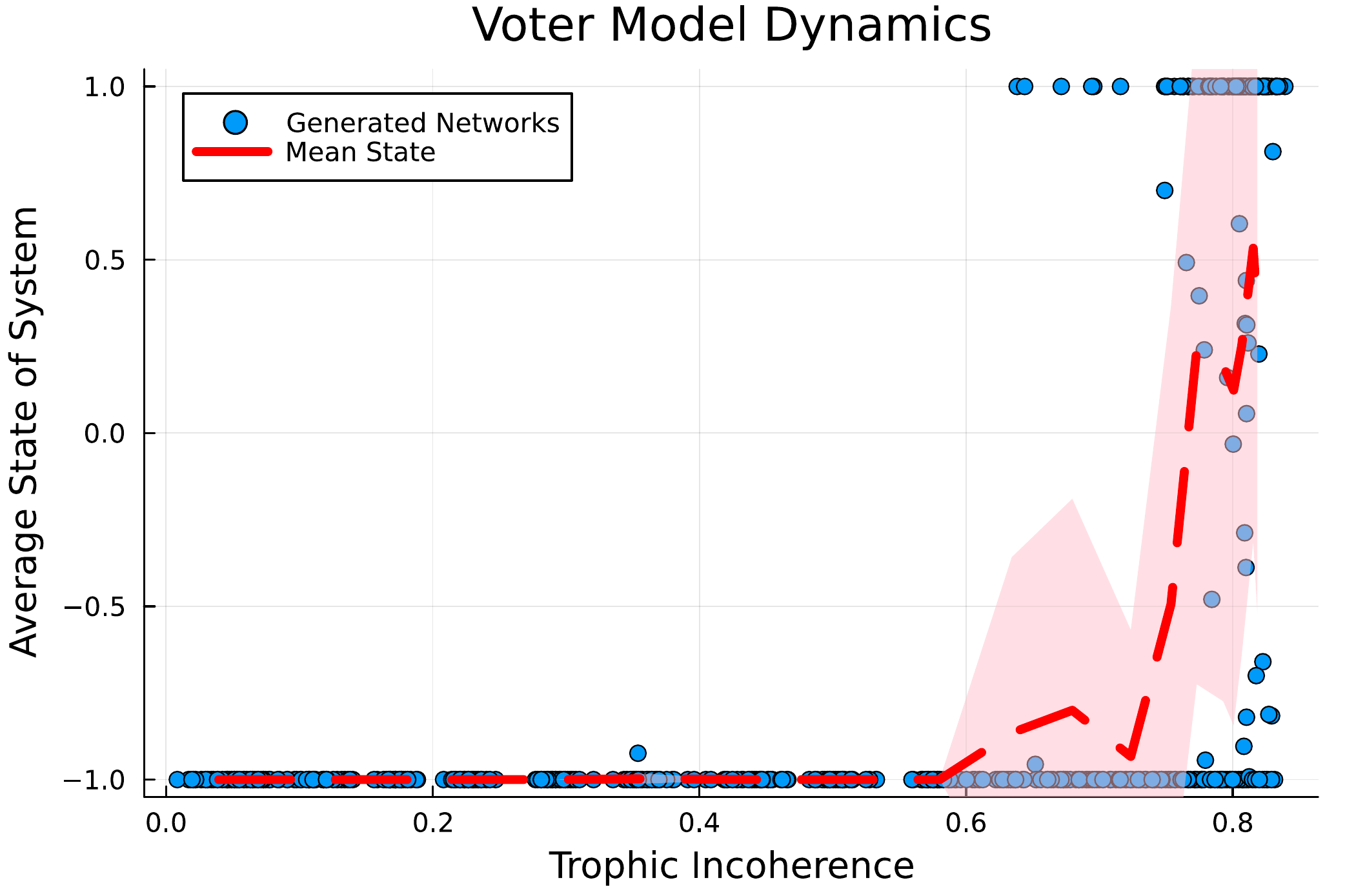}
         \caption{Influence by 20\% of nodes with lowest Trophic Level.}
         \label{fig:vote_20}
     \end{subfigure}
     \hfill
     \begin{subfigure}[t]{0.45\textwidth}
         \centering
         \includegraphics[width=\textwidth]{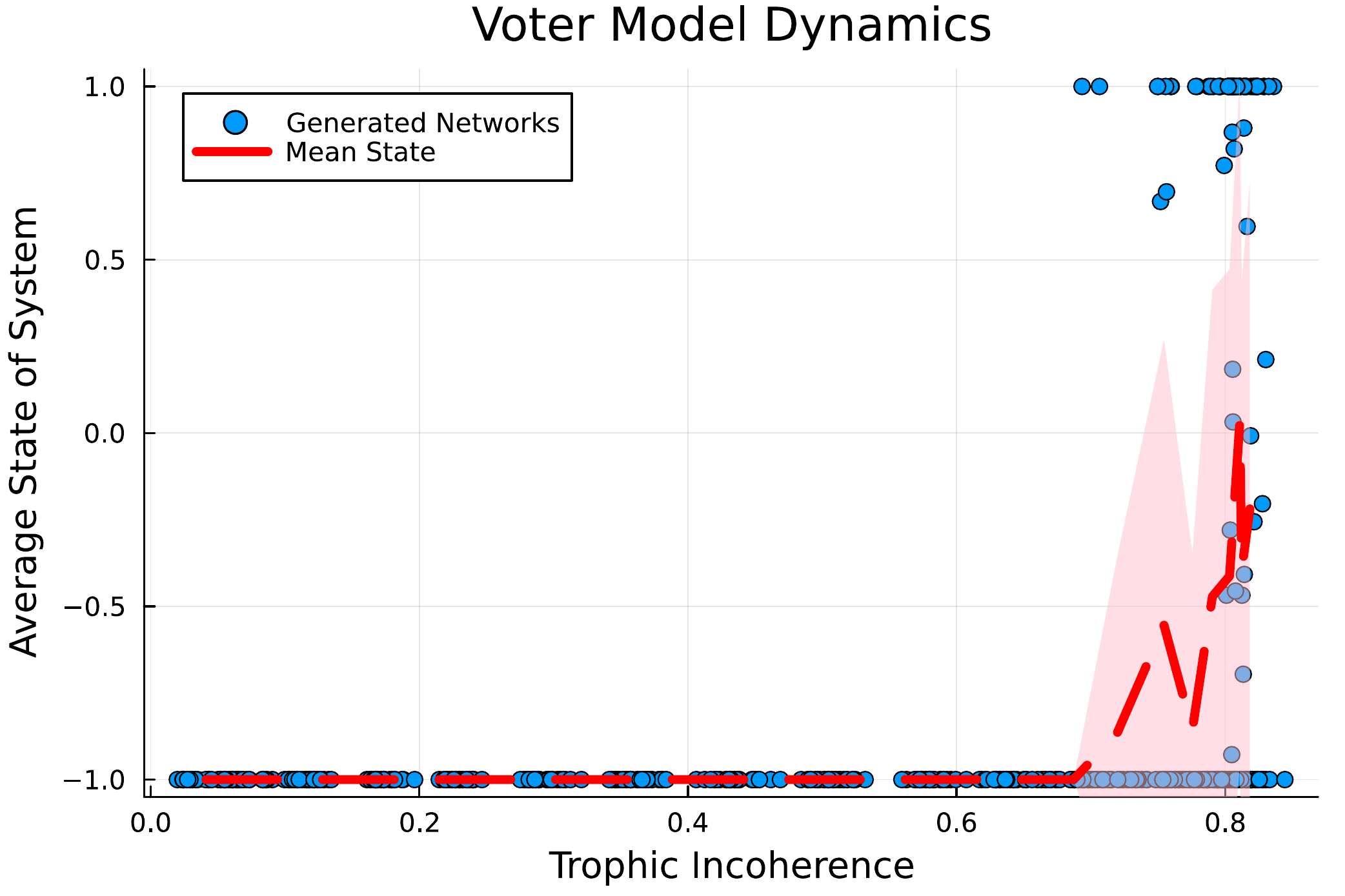}
         \caption{Influence by 40\% of nodes with lowest Trophic Level.}
         \label{fig:vote_40}
     \end{subfigure}
        \caption{Voter Model Dynamics \textcolor{black}{after 1000 updates} when different fractions nodes, chosen by lowest Trophic Level, are set to the initial condition $-1$ while the rest of the nodes are initially $+1$; for varying Trophic Incoherence, \textcolor{black}{where each point is a distinct} \textcolor{black}{network generated by the model in section \ref{Net_gen}} \textcolor{black}{with} $N=500$, $\langle k \rangle = 5$ \textcolor{black}{and} no basal nodes. \textcolor{black}{Temperatures, $T_{\text{Gen}}$, are logarithmically spaced between $10^{-1}$ and $10^2$ with 20 distinct temperatures used. At each temperature we generate 30 networks and compute the mean and standard deviation of this set, with the mean plotted with the red dashed line and the standard deviation shown with the shaded area.}}
        \label{fig:voter_multiple_examples}
\end{figure}

Similar results \textcolor{black}{to the previously obtained ones} can be seen, where a new opinion introduced at the lowest \textcolor{black}{trophic} level nodes spreads well in \textcolor{black}{coherent} networks
but is less likely to establish itself in
\textcolor{black}{more incoherent}
networks.
This is demonstrated in figure \ref{fig:voter_multiple_examples}. \textcolor{black}{In the low incoherence networks the standard deviation is quite small and the ability of the low level nodes to \textcolor{black}{influence} the system is clear.  As we increase incoherence this is no longer always the case and we pass through a region of higher variability as we transition to the incoherent regime, where the dynamics cannot be as easily \textcolor{black}{shaped} by the low level nodes. When 40\% of the nodes are \textcolor{black}{set to -1}, figure \ref{fig:vote_40}, almost all the networks flip to the minority state and only a small number of very incoherent networks maintain the majority state. Very similar behaviour is found when 20\% of the nodes are \textcolor{black}{set to -1}, figure \ref{fig:vote_20}, with most of the networks being easily influenced and only some of the high incoherence networks being resilient to the perturbation. When only 5\% are \textcolor{black}{set to -1}, figure \ref{fig:vote_5}, we see the same trend but a smaller fraction of the networks are \textcolor{black}{influenced}, with more networks being outliers from the trend. In a similar way to the majority rule case, targeting a very small number of nodes makes it more likely that an insufficient number of edges leave the perturbed set to influence the whole network
\cite{Rodgers2022NetworkNetworks}.}

\textcolor{black}{This section has shown how in different dynamical processes the nodes which can be regarded as influential can be determined by using trophic levels, and the ability of the network to be influenced is heavily shaped by the trophic coherence.}

\subsection{Generalised Rock-Paper-Scissors Dynamics}

\begin{figure}[H]
      		\centering
            \includegraphics[width=0.5\linewidth]{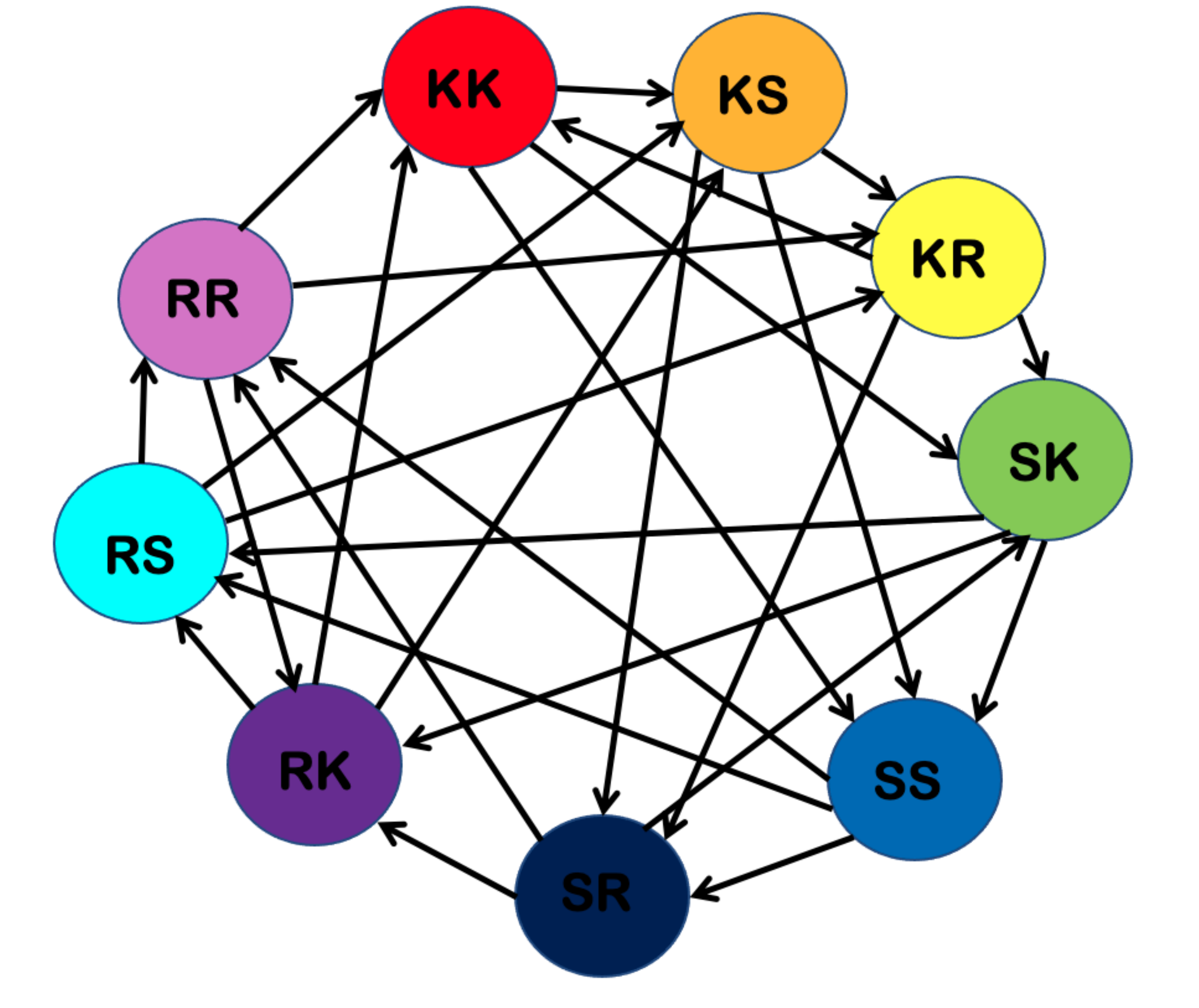}
        	\caption{\textcolor{black}{Example of a network of agents competing with each other through a complex interaction topology where an edge represents that an agent type is dominant over another. This is a network of bacteria competing through toxin warfare \cite{Szabo2007CompetingToxins}.
         Each node represents a type of bacteria (a strategy), which dominates over those types it has directed edges to, and is dominated by those it receives edges from. Hence, the network represents the relationship between bacteria rather than a spatial structure.}}

        \label{fig:bacetrial_network}
        \end{figure}

We show that Trophic level can also be used as an analysis tool for the states of generalised Rock-Paper-Scissors games where the interactions between strategies are defined by a complex network. \textcolor{black}{A standard Rock-Paper-Scissors game can be described with a directed cycle, in which each strategy has an edge to the strategy it defeats. In this case, there is no overall best choice of strategy thanks to the symmetry of the situation. This, however, may not be the case when the strategies interact through \textcolor{black}{a more complex ``strategy network'' which does not feature this cyclic structure (e.g. interactions between strategies and characters in video-games or ecological interactions such as competition between bacteria \cite{Szabo2007CompetingToxins}, as shown in figure \ref{fig:bacetrial_network}). In such cases there may be strategies} which perform better than others. }

\textcolor{black}{The interaction between individual agents} can be defined on a spatial structure like a lattice or by allowing every agent to interact with every other \textcolor{black}{as in a} complete graph. For simplicity we allow all the agents to interact with each other and ignore spatial structure. This setup shares many similarities with mean-field analysis of Lotka-Volterra systems, which \textcolor{black}{have} been studied with respect to trophic levels \textcolor{black}{and coherence} in 
\cite{Johnson2014TrophicStability}.
 \textcolor{black}{It has also} been much more widely studied in the regime of large random systems \cite{GarciaLorenzana2022Well-mixedInteractions,Mambuca2022DynamicalOscillations}. Most studies on non-random systems have looked at smaller systems, quite often with spatial aspects to them \cite{Szolnoki2014CyclicReview}. There has been some work looking at networks of generalised Rock-Paper-Scissors games but it mostly focuses on varying the strength of parameters rather than looking at large varied structures. This \textcolor{black}{work} has mostly been \textcolor{black}{numerical} \cite{Avelino2020PerformanceSpecies, Avelino2022Lotka-VolterraLink, Avelino2022ParityNS12,Milne2021WeakModels,Menezes2023HowModels}, with some looking at \textcolor{black}{the behaviour of smaller systems with weighted interactions and weaker species}. There has been some analytical work on \textcolor{black}{Rock-Paper-Scissors-Lizard-Spock networks which have a more complex topology than the basic Rock-Paper-Scissors game} \cite{Postlethwaite2022StabilityRockPaperScissorsLizardSpock}.

The setup used \textcolor{black}{here} is a graph of 100 nodes representing the possible strategies.
Then 1000 players per graph play each other at random with the interactions determined by the strategy graph.
\textcolor{black}{The initial strategy of each node is randomly assigned from the set of 100 possible strategies.}
If an edge points from strategy $i$ to strategy $j$ then if a player playing strategy $j$ meets a player playing strategy $i$ they will be beaten and \textcolor{black}{switch from strategy $i$} to strategy $j$. This definition could easily be flipped and then successful strategies would be those of high trophic level. However, we use this convention to match the earlier results. \textcolor{black}{These} dynamics can be quite complicated to analyse for several reasons. `Weaker' strategies may survive better than expected if the strategies that they are weak to are made extinct before them. The presence of cycles and neutral interactions means that one strategy taking over everywhere is not the norm and instead the system moves to a dynamic equilibrium where many of the strategies coexist \textcolor{black}{and ``defensive alliances'' are possible \cite{Szabo2007CompetingToxins}}. What makes a strategy successful is \textcolor{black}{a complex problem - it depends on the balance of how many strategies it beats or is beaten by,} as well as where those strategies fit into the global meta-game. For example, it is better to be weak to a strategy which is overall not widely used and then strong against a strategy which is widely played.
In this setup a strategy which is influential is one which \textcolor{black}{becomes} widely played. As a test for the ability of trophic level to predict influence we compare the probability of a strategy being played with the trophic level ranking of a node and then compare this to ranking by traditional centrality metrics such as PageRank \cite{Google1999,Gleich2015PageRankWeb}.

\begin{figure}[H]
     \centering
     \begin{subfigure}[t]{0.45\textwidth}
         \centering
         \includegraphics[width=\textwidth]{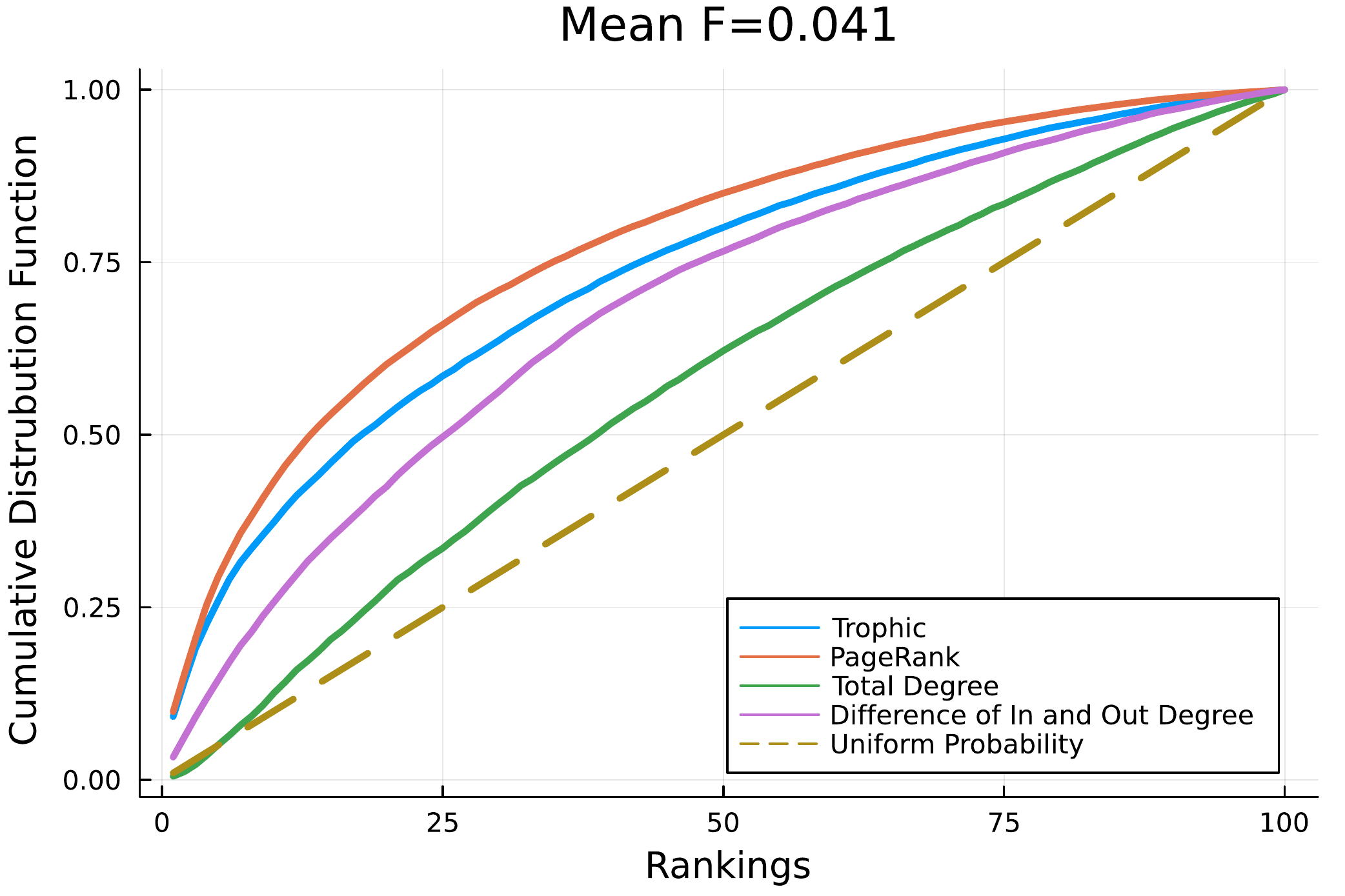}
         \caption{Networks of low trophic incoherence}
         \label{fig:RPS_low}
     \end{subfigure}
     \hfill
     \begin{subfigure}[t]{0.45\textwidth}
         \centering
         \includegraphics[width=\textwidth]{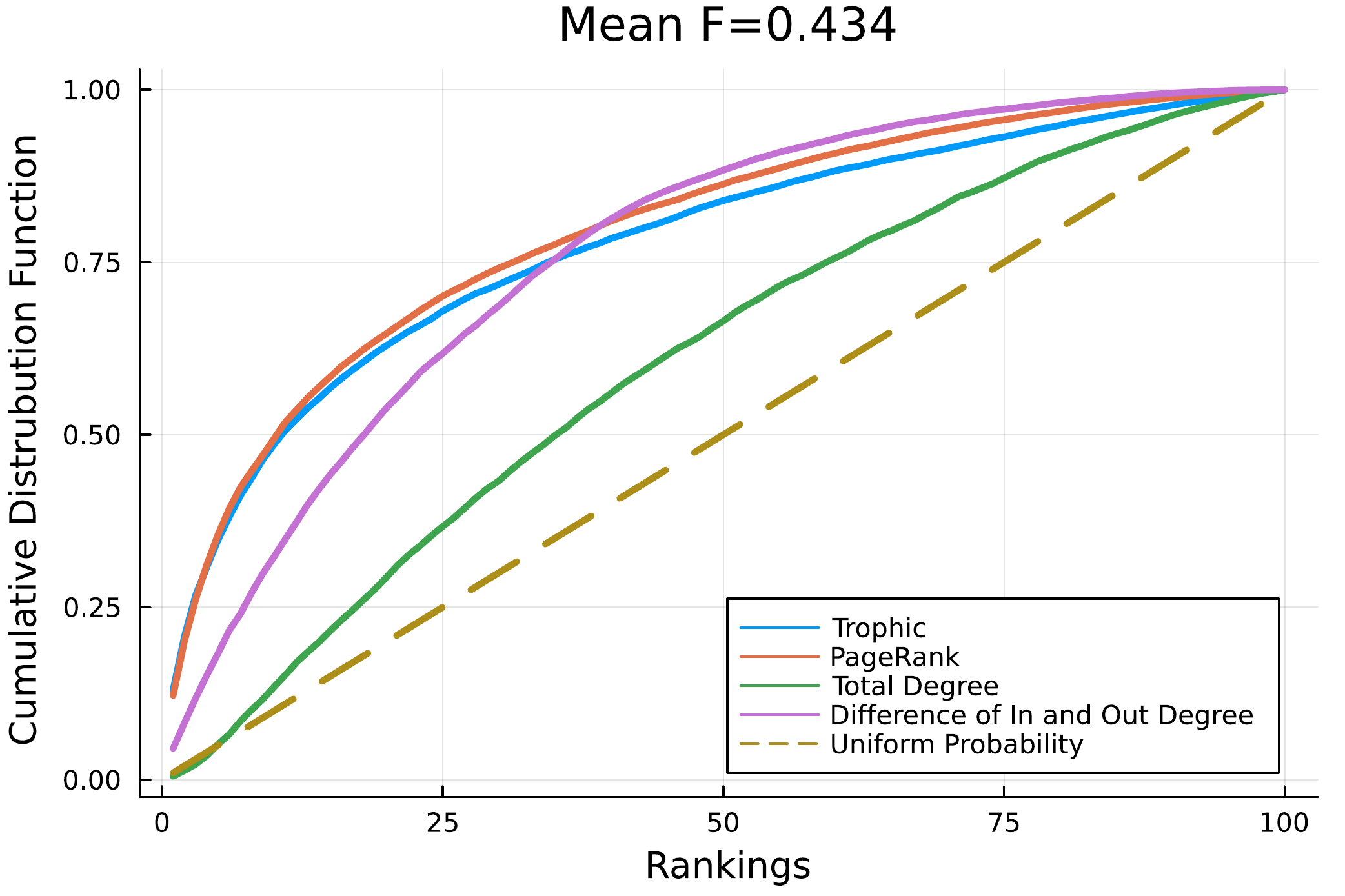}
         \caption{Networks of intermediate  trophic incoherence}
         \label{fig:RPS_med}
     \end{subfigure}
     \hfill
     \begin{subfigure}[t]{0.45\textwidth}
         \centering
         \includegraphics[width=\textwidth]{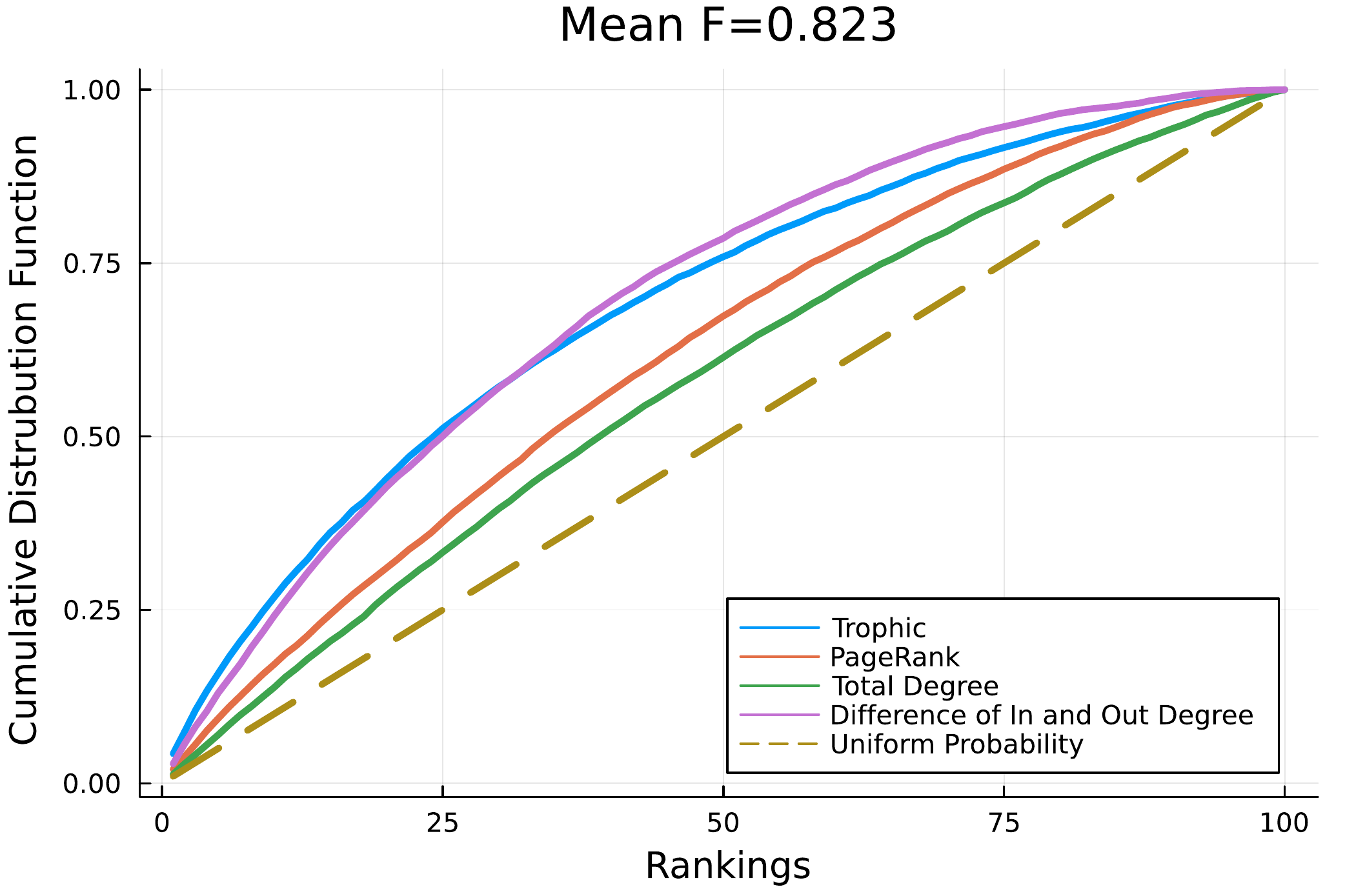}
         \caption{Networks of high  trophic incoherence}
         \label{fig:RPS_high}
     \end{subfigure}
        \caption{Cumulative Distributions of probability of playing a strategy per player per game per network by different centrality rankings and different mean incoherence over 1000 network  samples \textcolor{black}{ generated by the model in section \ref{Net_gen} at fixed temperatures corresponding to low, medium and high incoherence: $T_{\text{GEN}} = 0.02,1$   \textcolor{black}{and $100$, respectively. Each network has}} with $N=100$, \textcolor{black}{$\langle k \rangle =5$} and 1000 players playing a generalised Rock-Paper-Scissors game.}
        \label{fig:RPS_examples}
\end{figure}

\textcolor{black}{The results are shown in figure \ref{fig:RPS_examples}} for \textcolor{black}{networks with low, intermediate and high trophic incoherence.}
\textcolor{black}{
We consider the following centralities: trophic level, PageRank, out-degree, and imbalance (the difference between in-degree and out-degree). In the case of PageRank and imbalance, we apply the centrality to the adjacency matrix such that influence should correlate with a high out-degree, which may require transposing the matrix depending on convention used. The higher the area under the curve for a given centrality, the better it performs at identifying the most successful strategies. 
}

For the most coherent networks, \textcolor{black}{the best predictor of influence is PageRank, closely followed by trophic level, }figure \ref{fig:RPS_low}. \textcolor{black}{In these networks PageRank correlates with trophic level but also reveals information about a node's reach, and we hypothesise that both these features contribute to a strategy's success.} 
For networks of intermediate incoherence both trophic level and PageRank again are the best predictors of a strategy being successful, figure \ref{fig:RPS_med}\textcolor{black}{, although imbalance is also now a good measure}. However, this changes for networks with \textcolor{black}{high incoherence,} where the imbalance between \textcolor{black}{in- and out-degrees} starts to matter more \textcolor{black}{than PageRank.}  Trophic level and the degree imbalance are \textcolor{black}{now the best predictors, with PageRank performing similarly to out-degree}, figure \ref{fig:RPS_high}. This demonstrates how trophic level is a good measure of influence as it compares well to the best centrality measure in each case. It also highlights the utility of trophic coherence as a measure to understand network structure, since it \textcolor{black}{helps to explain} why the best centrality measure \textcolor{black}{depends on the network}.
Similar results can be found for many standard network types like random graphs or scale-free \textcolor{black}{networks, }where again trophic level is competitive with the best other centrality \textcolor{black}{metrics}. \textcolor{black}{It is noteworthy that trophic level is a good predictor of strategy success even in highly incoherent networks. However, if the network were maximally incoherent ($F=1$), all the nodes would have the same trophic level and hence, according to this predictor, an equal probability of success. This is the case, for instance, in the standard Rock-Paper-Scissors game.}

\section{Influence and Influenceability of Structure }

\textcolor{black}{In this section we demonstrate how several topological notions of `influence' and `influenceability' based on the eigenvectors of the adjacency matrix can be understood and interpreted by looking at the trophic structure of directed networks. This section provides an explanation for the effect of trophic coherence on dynamics seen in the last section, and how  networks can be better understood through trophic analysis. In particular, we show why the importance of low trophic level nodes depends on the network's trophic coherence. We provide results for numerically generated networks to control for variations in size, as well as results on a diverse data-set of real-world networks \cite{DataSamJohnson}. \textcolor{black}{In real-world networks, such as the ones contained within our sample,} there are a variety of network sizes, mean degrees and degree distributions. \textcolor{black}{However, we show} in this section that the general trends hold in both the numerically generated and real-world networks.}

\subsection{Localisation of Eigenvectors}

One simple way to quantify influence in a network is eigenvector centrality, a widely used centrality metric \cite{Newman2010Networks:Introduction}. The eigenvector centrality score of a node is determined by the centrality scores of its neighbours \cite{Newman2010Networks:Introduction}. This makes sense as we assume that important nodes also connect to other nodes of high importance. This can be shown to be equivalent to the principal eigenvector of the adjacency matrix and provides one measure of the importance of nodes in a network \cite{Newman2010Networks:Introduction}. 
\textcolor{black}{As we shall see, in trophically coherent} networks the eigenvectors are localised. This has many implications for the dynamics and affects the sensitivity of the network to perturbation and being \textcolor{black}{disrupted} by a small subset of nodes. The localisation of eigenvectors was first discovered in physics \cite{Anderson1958AbsenceLattices} where the phenomenon of electron localisation known as Anderson localisation was put forward. It has more recently been widely studied in random matrices \cite{LucasMetz2019SpectralMatrices}. It has wide implications for the dynamics of both physical systems and biological neural networks \cite{Zhang2019EigenvalueMatrices, Amir2016Non-HermitianNetworks,Tanaka2019Non-HermitianNetworks}.

The localisation of the eigenvectors can be measured in a variety of ways and it is also important to consider which nodes the eigenvectors localise around. As the ability to influence a network depends on the distribution of the centrality and where the high centrality nodes are in the network.

\subsubsection{Inverse Participation Ratio}

\begin{figure}[H]
    \centering
    
\begin{subfigure}[t]{0.45\textwidth}	
            \centering
            \includegraphics[width=\textwidth]{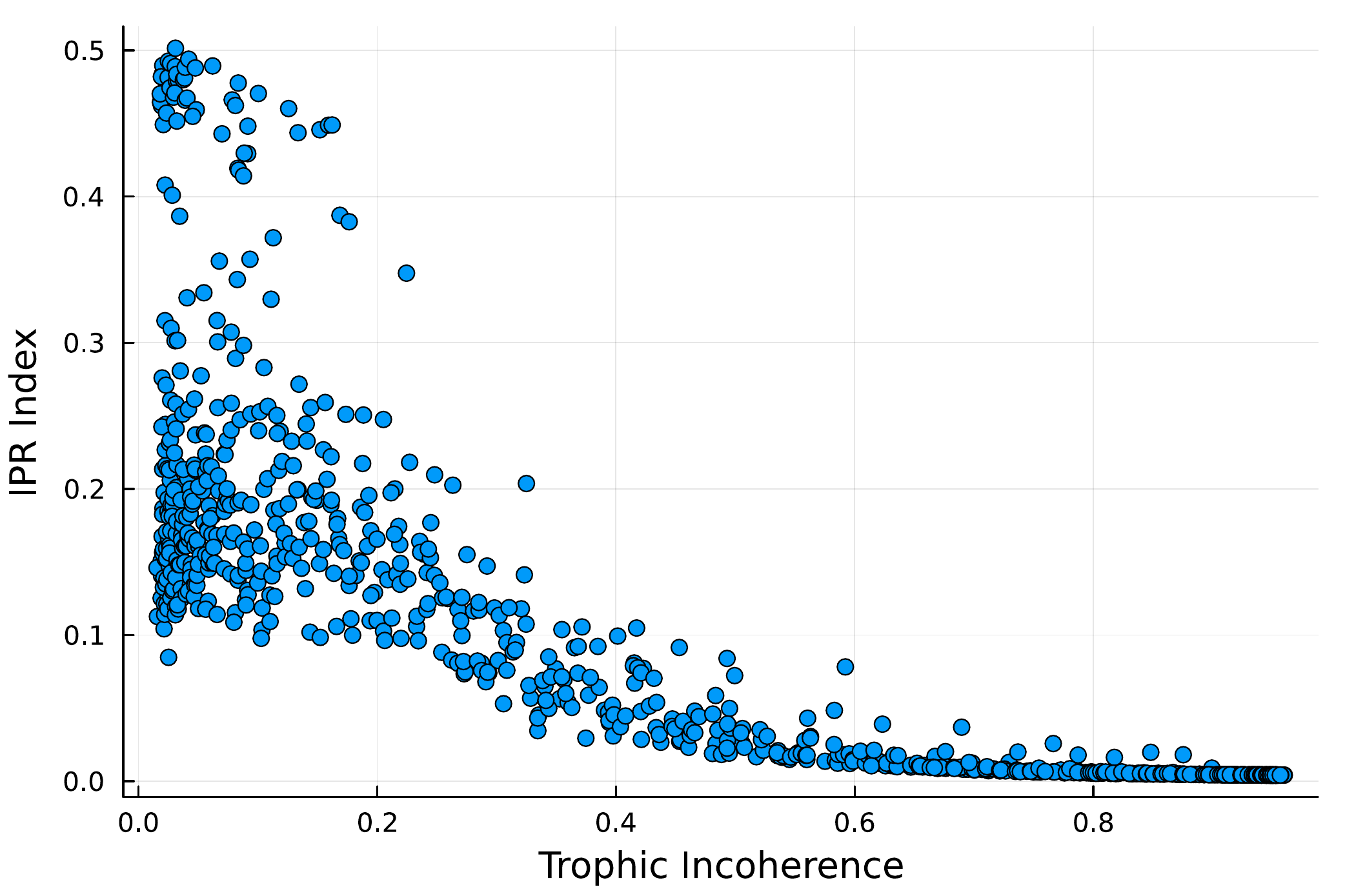}
        	\caption{\textcolor{black}{1000 data points each representing a \textcolor{black}{network generated by the model in section \ref{Net_gen}} of $N=500$ nodes, $\langle k \rangle =20$.
        	\textcolor{black}{Generation temperatures are logarithmically spaced between $10^{-2}$ and $10^2$.}}}
        \label{fig:eigen_IPR_average}
        \end{subfigure}
        \hfill
\begin{subfigure}[t]{0.45\textwidth}
            \centering
            \includegraphics[width=\textwidth]{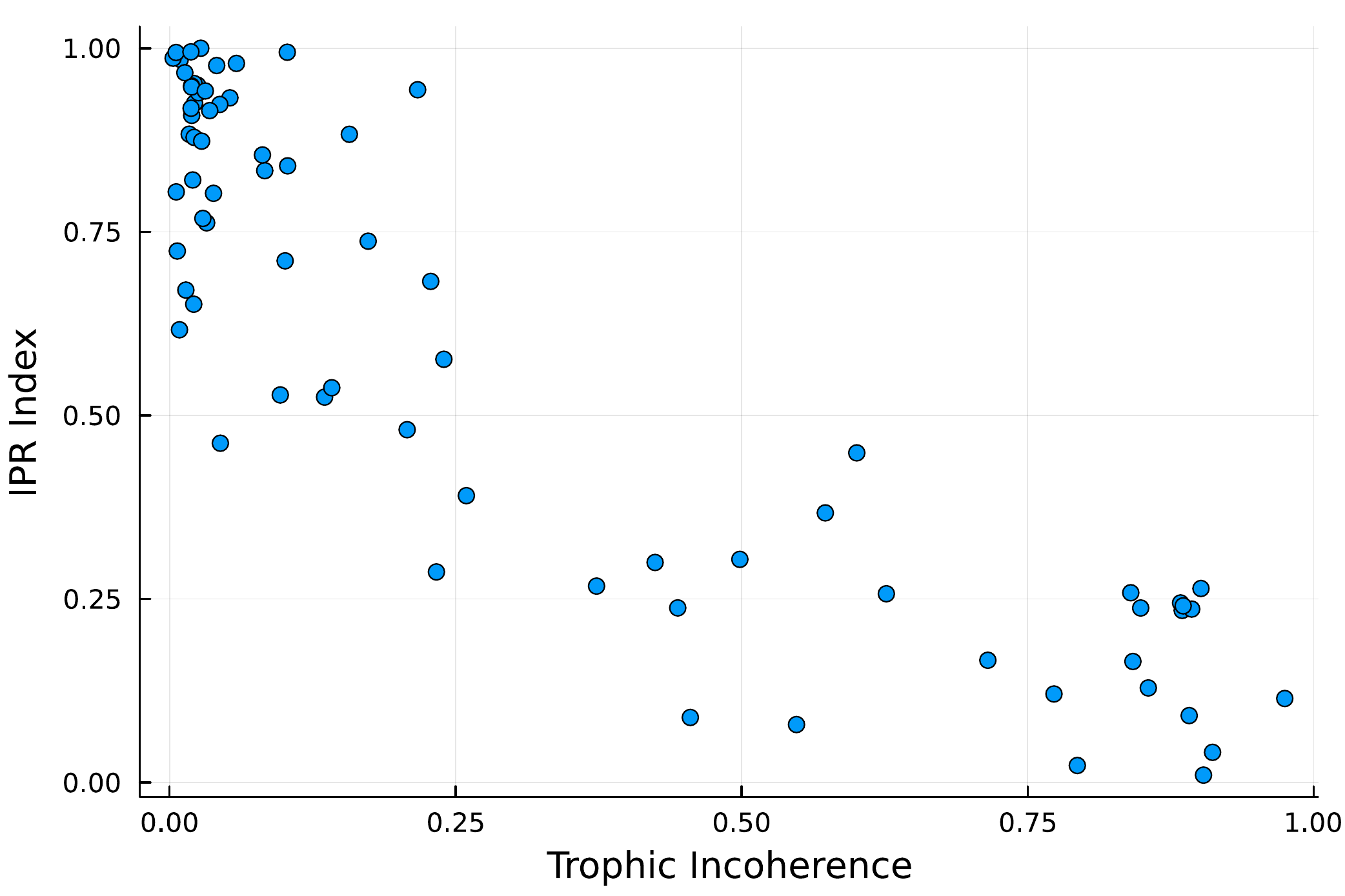}
        	\caption{\textcolor{black}{Real Networks- Taken from \cite{DataSamJohnson}. Each point representing a single network.}}
        	
        \label{fig:eigen_IPR_average_real}
    \end{subfigure}    
    \caption{\textcolor{black}{Average Inverse Participation Ratio (IPR) of $L^2$ normalised eigenvectors of the adjacency matrix of real and generated networks for varying trophic incoherence}}
    \label{fig:IPR_results}
\end{figure}

The inverse participation ratio measures the localisation of an eigenvector. It is defined for the $n\text{th}$ eigenvector as \begin{equation} \label{IPR}
    \text{IPR}_{n} = \frac{\sum_{i=1}^N \lvert \Psi^n_{i} \rvert ^4}{(\sum_{i=1}^N \lvert \Psi^n_{i} \rvert ^2)^2}.
\end{equation} Where $\lvert \Psi^n_{i} \rvert$ is the absolute value of each component of the \textcolor{black}{$n\text{th}$} eigenvector. This sum is maximised when all non-zero components are concentrated at one value and takes its minimum value when the non-zero components are exactly equally distributed.
\textcolor{black}{In figure \ref{fig:IPR_results} we show, for each network, the average $IPR_n$ over eigenvectors, $IPR=N^{-1}\sum_n IPR_n$. The localisation varies with Trophic Incoherence: it is large when the network is more coherent and the eigenvalues are localised; it decreases when the network is more incoherent and the eigenvectors delocalise. Moreover, at low incoherence there is a higher variability in $IPR$ between different networks.
}

\textcolor{black}{For numerically generated networks, \ref{fig:eigen_IPR_average}, the effect is very clear.} The inverse participation also follows the same trend in a \textcolor{black}{data-set of real world networks}, figure \ref{fig:eigen_IPR_average_real}, which includes food-webs, trade networks, neural and social networks \cite{DataSamJohnson}. \textcolor{black}{However, the real networks vary in size, density and degree heterogeneity, and thus present a noisier picture. The maximum localisation is also larger than in the numerically generated case.} \textcolor{black}{This helps to explain why in the previous section we found that a small number of low trophic level nodes can \textcolor{black}{influence} the dynamics when the network is more coherent, as this is the regime where the eigenvectors localise and `influence' is concentrated.}

\subsubsection{Entropy of Eigenvectors}

\begin{figure}[H]
    \centering
\begin{subfigure}[t]{0.45\textwidth}	
      		\centering
            \includegraphics[width=\textwidth]{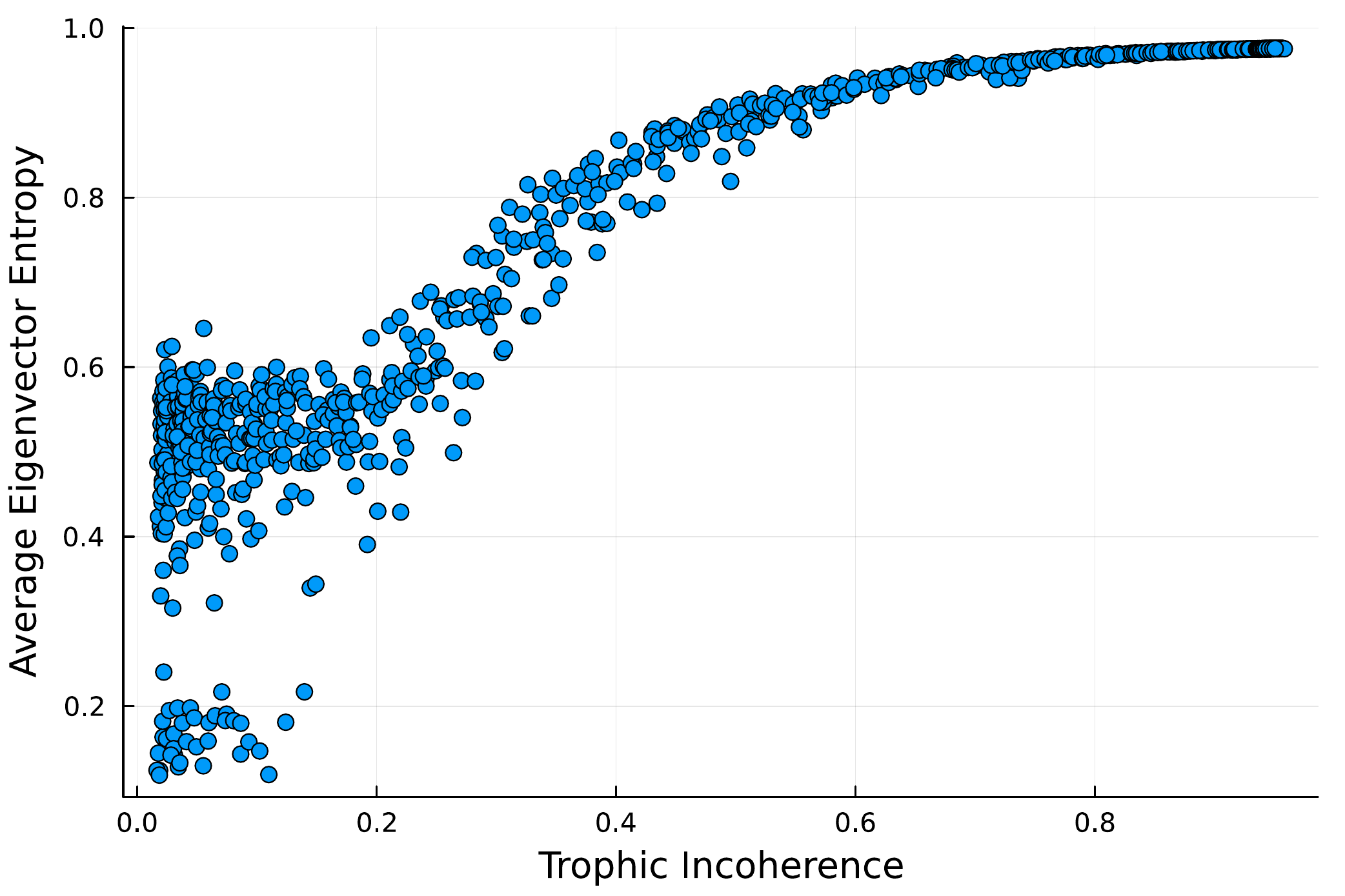}
        	\caption{\textcolor{black}{1000 data points each representing a \textcolor{black}{network generated by the model in section \ref{Net_gen}} of $N=500$ nodes, $\langle k \rangle =20$. \textcolor{black}{Generation temperatures are logarithmically spaced between $10^{-2}$ and $10^2$.}
        	}}
        	
        \label{fig:eigen_ENT_average}
        \end{subfigure}
        \hfill
\begin{subfigure}[t]{0.45\textwidth}	
      		\centering
            \includegraphics[width=\textwidth]{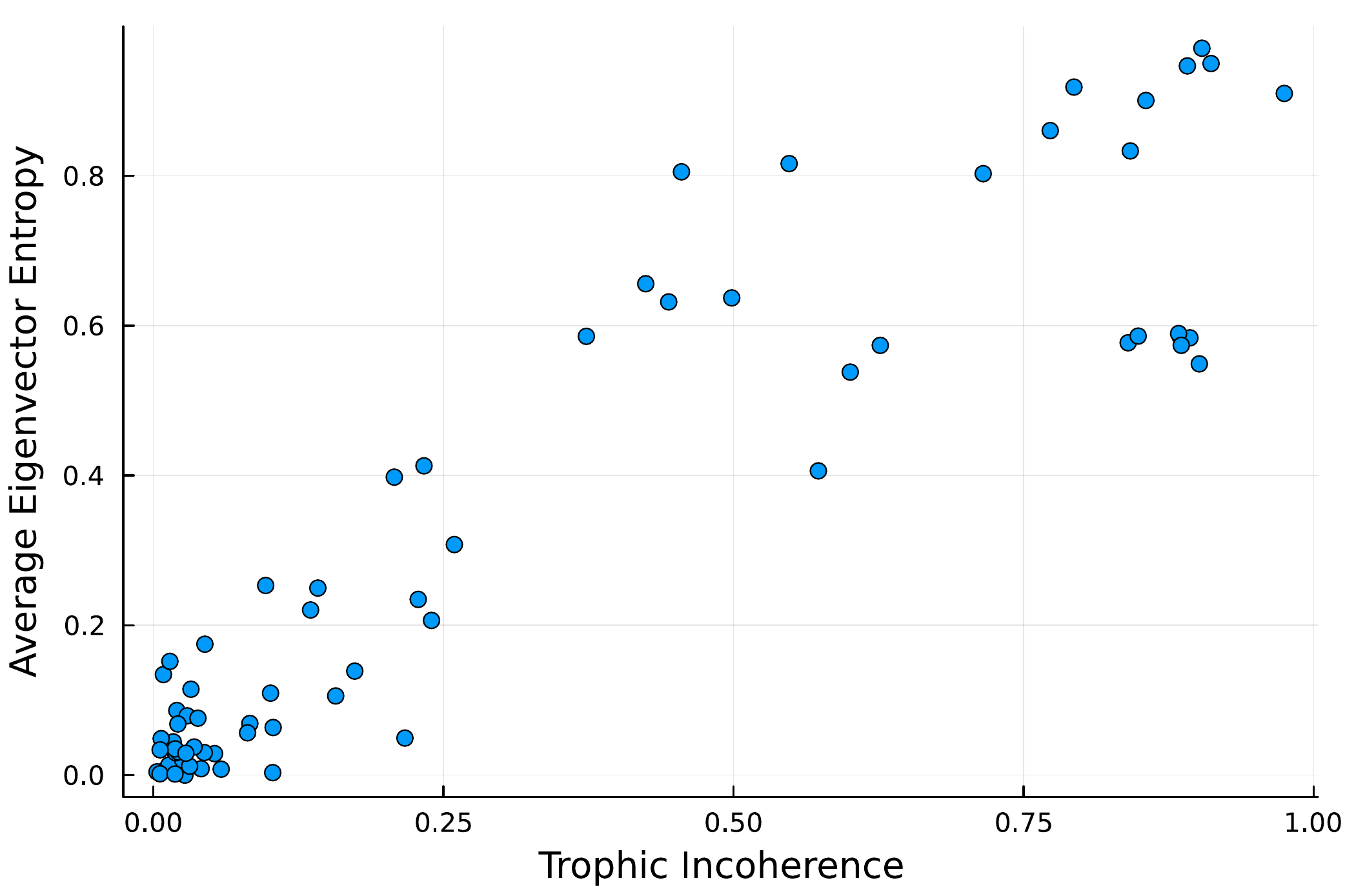}
        	\caption{\textcolor{black}{Real networks from \cite{DataSamJohnson}. Each point represents a single network.} }
        	
        \label{fig:eigen_ENT_real}
        \end{subfigure}
   \caption{ \textcolor{black}{Average entropy of $L^1$ normalised eigenvectors of the adjacency matrix of real and generated networks for varying trophic incoherence}}
    \label{fig:ent_results}
\end{figure}

The entropy of an eigenvector is another way to measure localisation based on the information about the system carried by the vector. It is defined \textcolor{black}{for the $n\text{th}$ eigenvector $\Phi^n$ (normalised so that the absolute values of the elements sum to 1)} as \begin{equation} \label{eq:entropy}
    S(\Phi^n)  = \frac{-1}{\ln{N}}\sum_{i=1}^N  \lvert \Phi^n_{i} \rvert \ln{\lvert \Phi^n_{i} \rvert}.
\end{equation}   
This is minimised when all the weight of the eigenvector is concentrated in one node, and hence we know the most information about the system as all the importance is concentrated in that node. \textcolor{black}{The entropy is} maximised when the weight of all nodes is equal and all nodes are equivalent, so we have the least \textcolor{black}{knowledge} about the structure of the system. \textcolor{black}{We normalise by the maximum value, $\ln N$, so that that $0\leq S\leq 1$ and we can compare networks of different size. A network can be characterised by the average, $\bar{S}=N^{-1}\sum_n S(\Phi^n)$} 
\textcolor{black}{In figure \ref{fig:ent_results} we show $\bar{S}$ for networks of varying trophic incoherence.}

\textcolor{black}{The trend in the numerically generated networks, figure \ref{fig:eigen_ENT_average}, is very clear, with the entropy of the very coherent networks being low and then increasing with incoherence.} \textcolor{black}{The real networks, figure \ref{fig:eigen_ENT_real}, produce a similar trend to the numerically generated networks, but again the data is noisier because of the more varied structure of real systems.} \textcolor{black}{This again is consistent with networks being sensitive to perturbations or attacks targeting the nodes of low trophic level, but the degree of sensitivity depending on the trophic coherence of the network.}

\subsection{Correlation Between Left and Right Eigenvectors}

\begin{figure}[H]
    \centering
\begin{subfigure}[t]{0.45\textwidth}	
      		\centering
            \includegraphics[width=\textwidth]{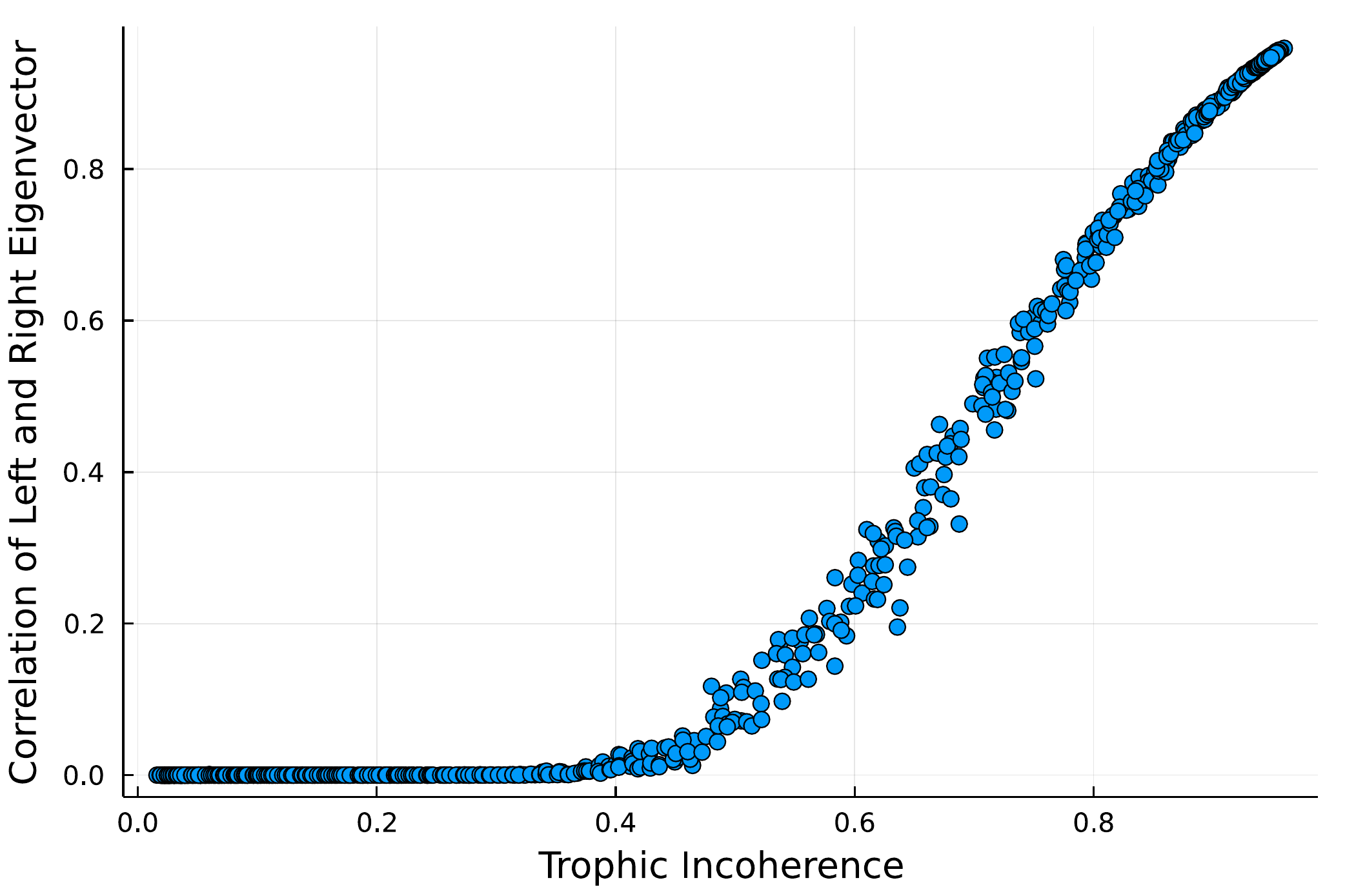}
        	\caption{\textcolor{black}{1000 data points each representing a \textcolor{black}{network generated by the model in section \ref{Net_gen}} of $N=500$ nodes, $\langle k \rangle =20$. \textcolor{black}{Generation temperatures are logarithmically spaced between $10^{-2}$ and $10^2$.}
        	}}
        	
        \label{fig:gen_networks_eigen_correlation}
        \end{subfigure}
        \hfill
\begin{subfigure}[t]{0.45\textwidth}	
      		\centering
            \includegraphics[width=\textwidth]{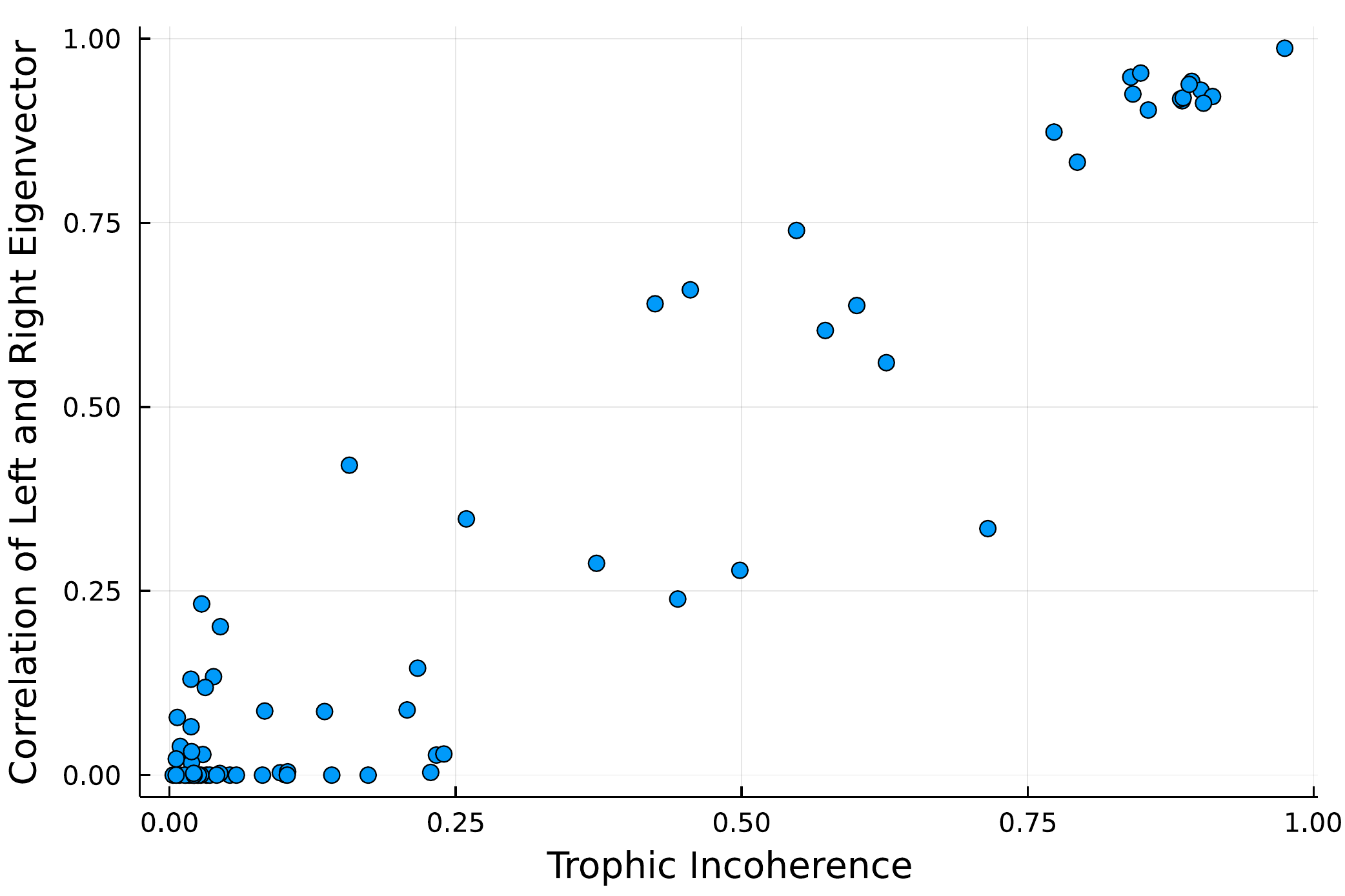}
        	\caption{\textcolor{black}{Real networks from \cite{DataSamJohnson}. Each point represents a single network.}  
        	}
        	
        \label{fig:Real_networks_eigen_correlation}
        \end{subfigure} 
        
 \caption{\textcolor{black}{Correlation between left and right principal eigenvectors of the adjacency matrix of generated (a) and real (b) networks for varying trophic incoherence. Eigenvectors are $L^2$ normalised.}}
    \label{fig:my_label_3}
\end{figure}

A node can be considered important if it is able to receive \textcolor{black}{information from} and input information \textcolor{black}{into} the rest of the network. In directed networks the underlying hierarchy \textcolor{black}{of trophic levels} may mean that the nodes which are able to reach the most nodes in the network and the nodes which are reached by the most in the network are distinct groups. This is different from the undirected case where a hub node can effectively do both. This can be understood by looking at the correlation between the left and right principal eigenvectors. The left eigenvector can be thought of as a measure of the centrality of a node based on in-degree and its ability to receive information, and the right eigenvector can be thought of as the ability to emit information. This is just convention and transposing the adjacency matrix swaps these roles. One way to understand this is to look at the scalar product between the eigenvector centrality, \textcolor{black}{given by the principal eigenvector} of the adjacency matrix, and \textcolor{black}{that of} its transpose. Due to the Perron–Frobenius theorem both of these eigenvectors are real and \textcolor{black}{non-negative} so the multiplication of each element of the vector is real and \textcolor{black}{non-negative}. This is shown in figures \ref{fig:gen_networks_eigen_correlation} and \ref{fig:Real_networks_eigen_correlation}. \textcolor{black}{We see that the correlation} is small when the networks are coherent, as nodes of high centrality do not overlap; while when the network does not have \textcolor{black}{such a well-defined} hierarchical structure \textcolor{black}{(i.e. it is more incoherent)} there is a larger overlap, \textcolor{black}{tending} towards 1 as the eigenvectors are both normalised. This highlights the effects of hierarchy on the notion of node importance and that hierarchy can induce an asymmetry in node behaviour.

\textcolor{black}{In the generated networks the correlation between the eigenvectors follows a smooth trend whilst again in the case of real networks the overall correlation is reproduced but the results are much more noisy.}
\textcolor{black}{These results} can also be viewed in terms of matrix normality, which measures how well the adjacency matrix commutes with its transpose. This has previously been shown to be linked to trophic incoherence \cite{Johnson2020DigraphsSystems,MacKay2020HowNetwork}. However,
\textcolor{black}{the correlation between right and left principal eigenvectors provides a more direct measure of whether the influential nodes are also those that are influenceable.}

\subsubsection{Example Networks}

\begin{figure}[H]
     \centering
     \begin{subfigure}[t]{0.45\textwidth}
         \centering
         \includegraphics[width=\textwidth]{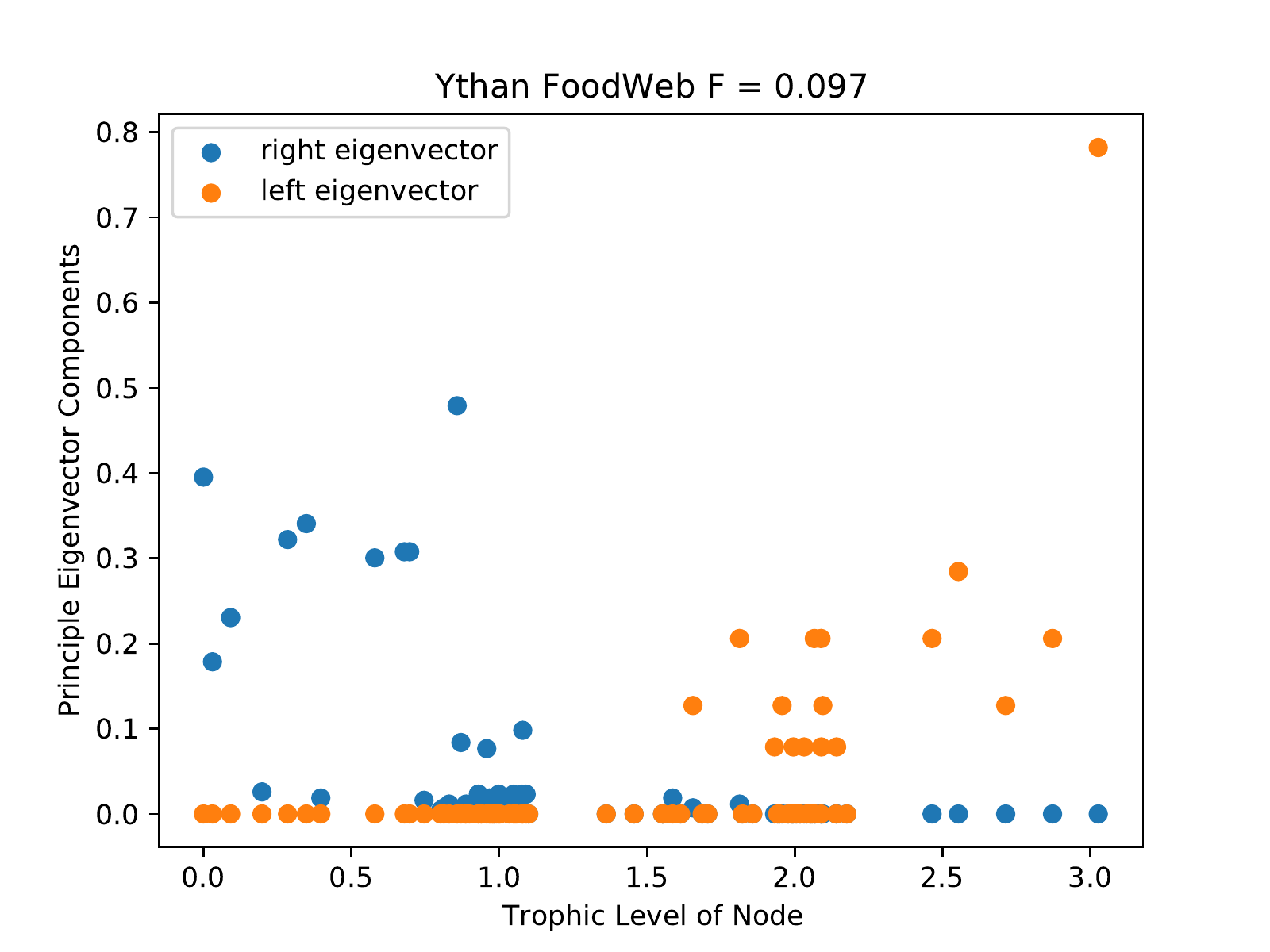}
         \caption{Ythan Estuary Food Web with low incoherence, $F=0.097$, \cite{DataSamJohnson}. \textcolor{black}{Scalar product between left and right eigenvectors 0.003.}}
         \label{fig:Ythan_eig}
     \end{subfigure}
     \hfill
     \begin{subfigure}[t]{0.45\textwidth}
         \centering
         \includegraphics[width=\textwidth]{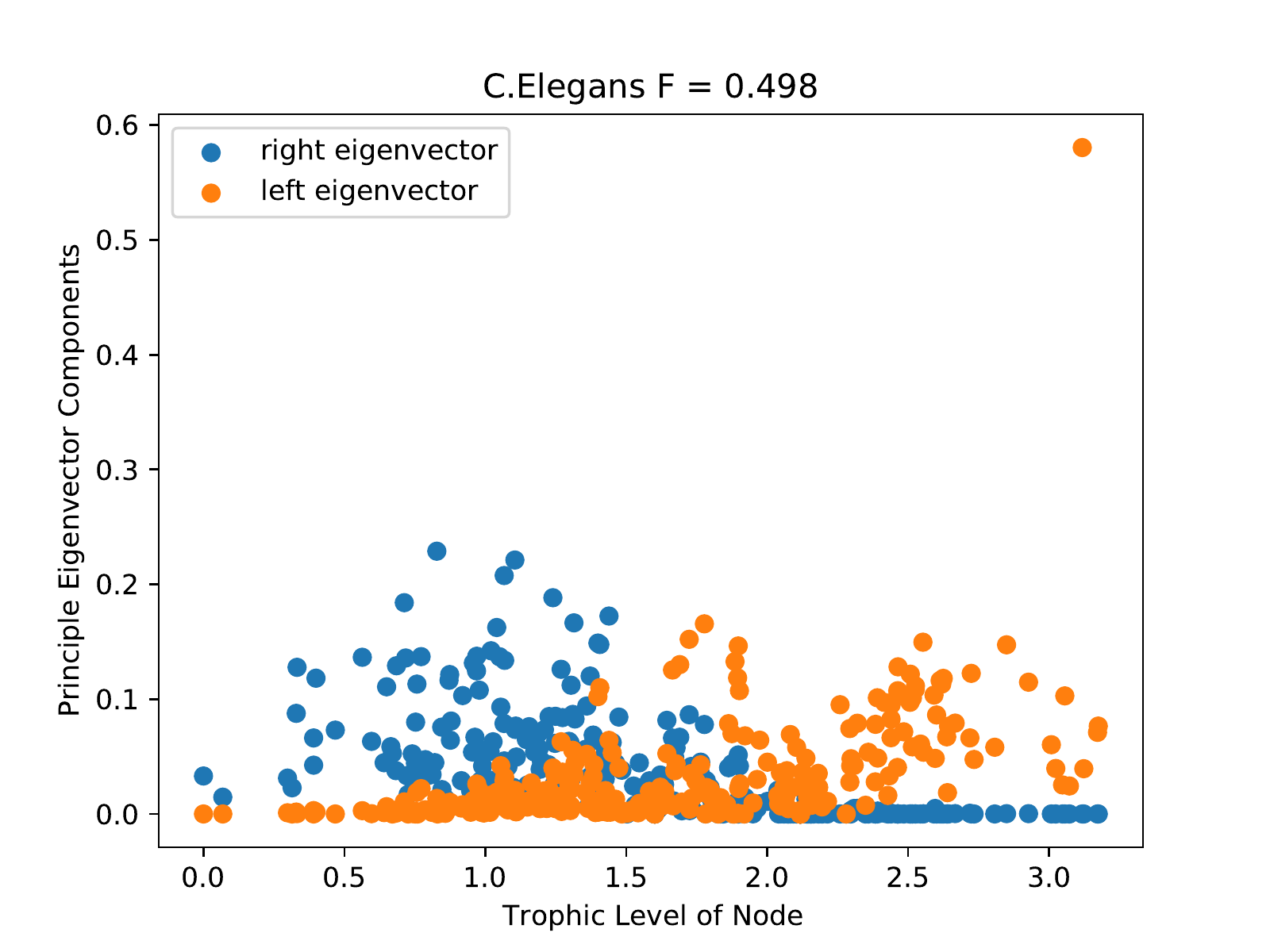}
         \caption{\textit{C.Elegans} Connectome of intermediate incoherence, $F=0.498$, \cite{DataSamJohnson}. \textcolor{black}{Scalar product between left and right eigenvectors 0.278.}}
         \label{fig:C_elegans_eig}
     \end{subfigure}
     \hfill
     \begin{subfigure}[t]{0.45\textwidth}
         \centering
         \includegraphics[width=\textwidth]{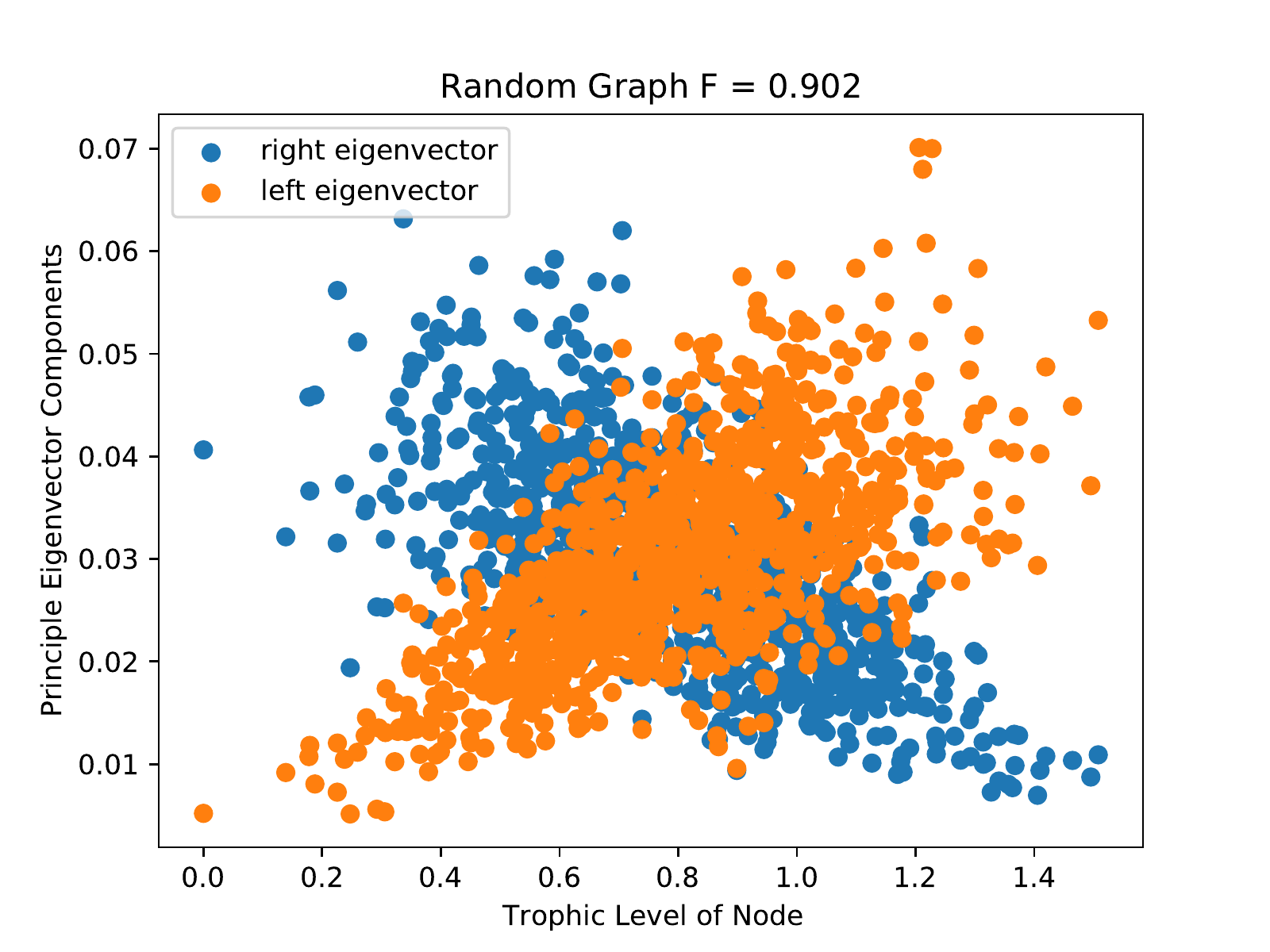}
         \caption{Random graph with $N=1000$, $\langle k \rangle =10 $ and high incoherence with $F=0.902$. \textcolor{black}{Scalar product between left and right eigenvectors 0.904.}}
         \label{fig:Random_eig}
     \end{subfigure}
       \hfill
     \begin{subfigure}[t]{0.45\textwidth}
         \centering
         \includegraphics[width=\textwidth]{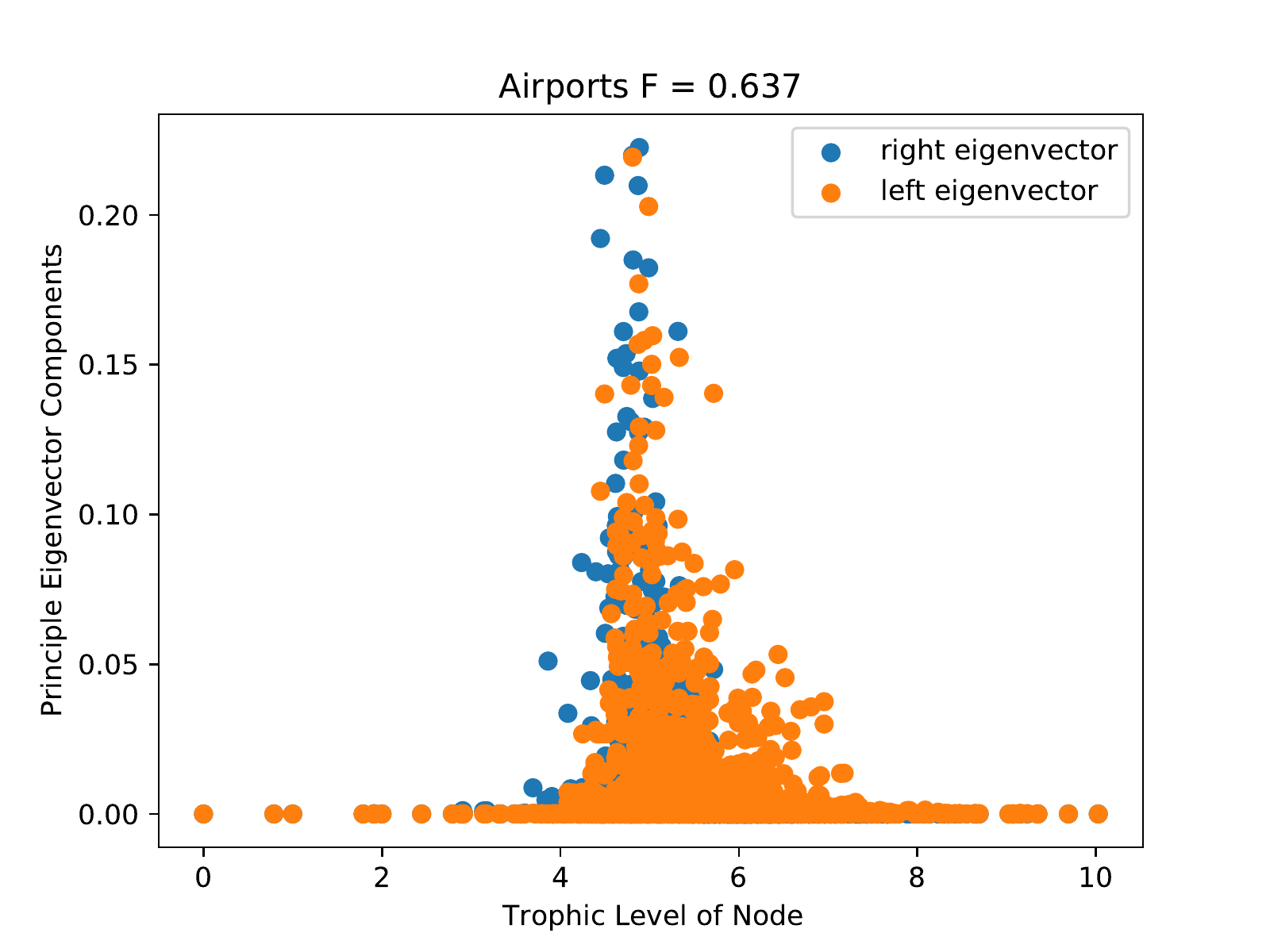}
         \caption{\textcolor{black}{FAA preferred routes between airports, $F=0.637$. Taken from  \cite{DataTiagoPeixotoairport}. Originally sourced from \cite{AirportData}. Scalar product between left and right eigenvectors 0.831.}}
         \label{fig:extra_eig}
     \end{subfigure}
     
        \caption{\textcolor{black}{Principal left and right eigenvectors ($L^2$ normalised) for four example networks, with the trophic level of the nodes on the horizontal axis.}}
        \label{fig:eig_exmaples}
\end{figure}

The previous results on the localisation of the eigenvectors and the correlation between the left and right eigenvectors, \textcolor{black}{on both numerically generated and real-world networks}, can be \textcolor{black}{better} understood by looking at some \textcolor{black}{specific} network examples. In coherent networks with a clear hierarchy \textcolor{black}{of trophic levels,} such as the Ythan estuary food web, figure \ref{fig:Ythan_eig}, the left and right eigenvectors localise to different parts of the network hierarchy. The nodes of large left eigenvector centrality are towards the top of the hierarchy while the nodes of large right eigenvector centrality centre on the nodes of lower trophic level. This reflects the intuition around the localisation of eigenvectors and the difference between the left and right eigenvectors, as well as what we know about food webs where energy flows from the bottom to the top of the system. This phenomenon can also be observed in real networks of intermediate coherence such as the connectome of the nematode \textit{C.Elegans}, figure \ref{fig:C_elegans_eig}. In this network the localisation is less pronounced than in the food web due to the network being more incoherent and having more feedback. There is, however, still clearly visible localisation with the non-zero elements not being evenly distributed with respect to trophic level, and the distribution of the non-zero elements of the left and right eigenvectors being centred at different places in the hierarchy.

A non-trivial relationship between trophic level and the distribution of eigenvectors can also be observed in \textcolor{black}{some} random graphs, figure \ref{fig:Random_eig}, where one would not expect there to be much useful hierarchical structure to observe. In a random graph the left and right eigenvectors are not as strongly localised and there are many non-zero elements shared by both the left and right eigenvectors. However, there is a positive correlation between the left eigenvector and trophic level, and a negative correlation between the right eigenvector and trophic level. This implies that nodes which are the best at emitting information and reaching the whole network are at the bottom of the hierarchy, nodes which receive information are at the top and nodes which do both equally are in the middle. This agrees with the well-known bow tie structure observed in directed networks \cite{Timar2017MappingDiagram}. 

\textcolor{black}{There do, however, exist networks which break this trend. Figure \ref{fig:extra_eig} shows the Federal Aviation Administration (FFA) preferred routes between airports. This network is directed, but since it is an airport network it features hubs which are hubs of both in and out degree. Due to the fact that certain airports are very important hubs in both senses, 
the left and right eigenvector peaks 
overlap
more than in previous real-world examples, as we can see in \ref{fig:extra_eig}.
Thus, this kind of network is quite different from, say, a food web, which has an overall directionality.
The airport network is more coherent than a random graph (in part because the degree imbalance is larger).
However, the level distribution does stretch out beyond the central peak to span a number of levels. This shows how trophic analysis can be used to identify 
structural features
related to network function.
In particular, 
it distinguishes clearly between
networks with a directional flow, like a food-web or neural network, 
and those lacking this property,
like the airport network.}

\subsection{Pseudospectra and Pseudospectral Radius}

A network can be influenced if it is sensitive to perturbations; \textcolor{black}{in particular,} if a disruption to a small number of nodes leads to a large 
\textcolor{black}{change in the behaviour of the system.}
This kind of phenomenon can be linked to the \textcolor{black}{non-normality of the adjacency matrix \cite{Trefethen2020SpectraPseudospectra}, i.e. how far it is from commuting} with its transpose.   
 \textcolor{black}{Networks which are highly} non-normal \textcolor{black}{(a feature that has been shown analytically to be related to} low trophic incoherence \cite{Johnson2020DigraphsSystems,MacKay2020HowNetwork}) have the property that \textcolor{black}{their} eigenvalues are extremely sensitive to perturbations \cite{Trefethen2020SpectraPseudospectra}. This can be impactful in many fields, such as fluid mechanics, acoustics, condensed matter physics and construction of numerical methods \cite{Trefethen2020SpectraPseudospectra, Okuma2020HermitianPseudospectra, Gebhardt1994ChaosStability, Symon2018Non-normalityAnalysis, Sujith2016Non-normalityInstabilities}.

 The pseudospectrum of a matrix $A$ with eigenspectrum $\sigma(A) $ is defined as \begin{equation}
    \sigma_{\epsilon}(A) = \{ \sigma(A+E): ||E|| \leq \epsilon \}
\end{equation} 
for some $\epsilon > 0$ and \textcolor{black}{$E$ a matrix of norm less than $\epsilon$.} It measures how the spectrum changes subject to perturbations, and can be computed using the Julia package \cite{Pseuodspectra.jl} based on EigTool \cite{EigTool}.
 The pseudospectra has recently been shown to be important in complex networks and the stability of ecosystems \cite{Asllani2018StructureNetworks}.

\begin{figure}[H]
     \centering
\begin{subfigure}[t]{0.45\textwidth}
      		\centering
            \includegraphics[width=\textwidth]{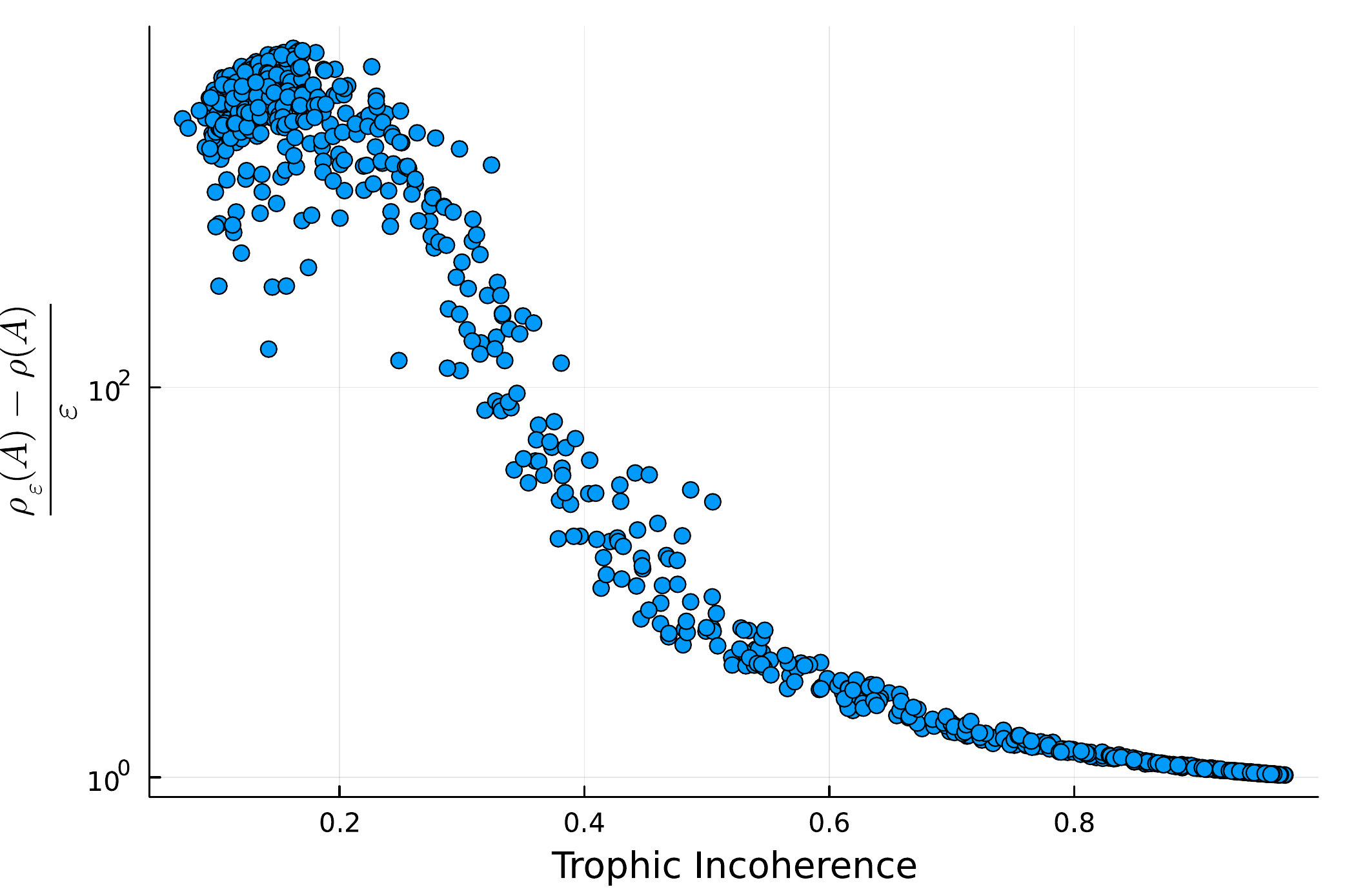}
        	\caption{  
        	\textcolor{black}{1000 data points each representing a \textcolor{black}{network generated by the model in section \ref{Net_gen}} of $N=100$ nodes, $\langle k \rangle =20$. Spectral perturbation size $10^{-3}$. \textcolor{black}{Generation temperatures are logarithmically spaced between $10^{-2}$ and $10^2$.} }}
        	
        \label{fig:pseudospectral_radius_100}
        \end{subfigure} 
        \hfill
\begin{subfigure}[t]{0.45\textwidth}
      		\centering
            \includegraphics[width=\textwidth]{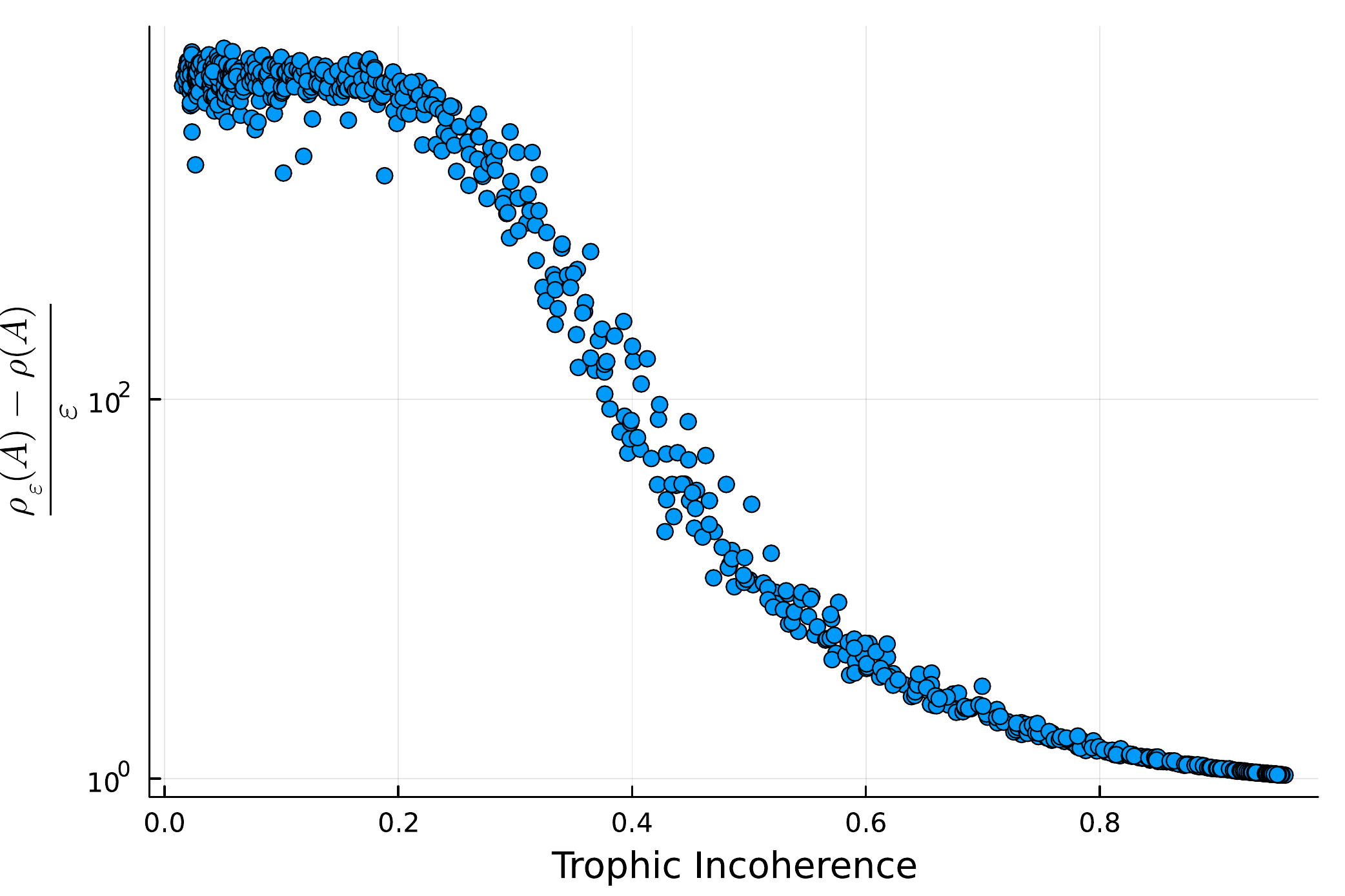}
        	\caption{ 
        	\textcolor{black}{1000 data points each representing a \textcolor{black}{network generated by the model in section \ref{Net_gen}} of $N=500$ nodes, $\langle k \rangle =20$. Spectral perturbation size $10^{-3}$. \textcolor{black}{Generation temperatures are logarithmically spaced between $10^{-2}$ and $10^2$.} }}
        \label{fig:pseudospectral_radius_500}
        \end{subfigure} 
        \hfill
\begin{subfigure}[t]{0.45\textwidth}
      		\centering
            \includegraphics[width=\textwidth]{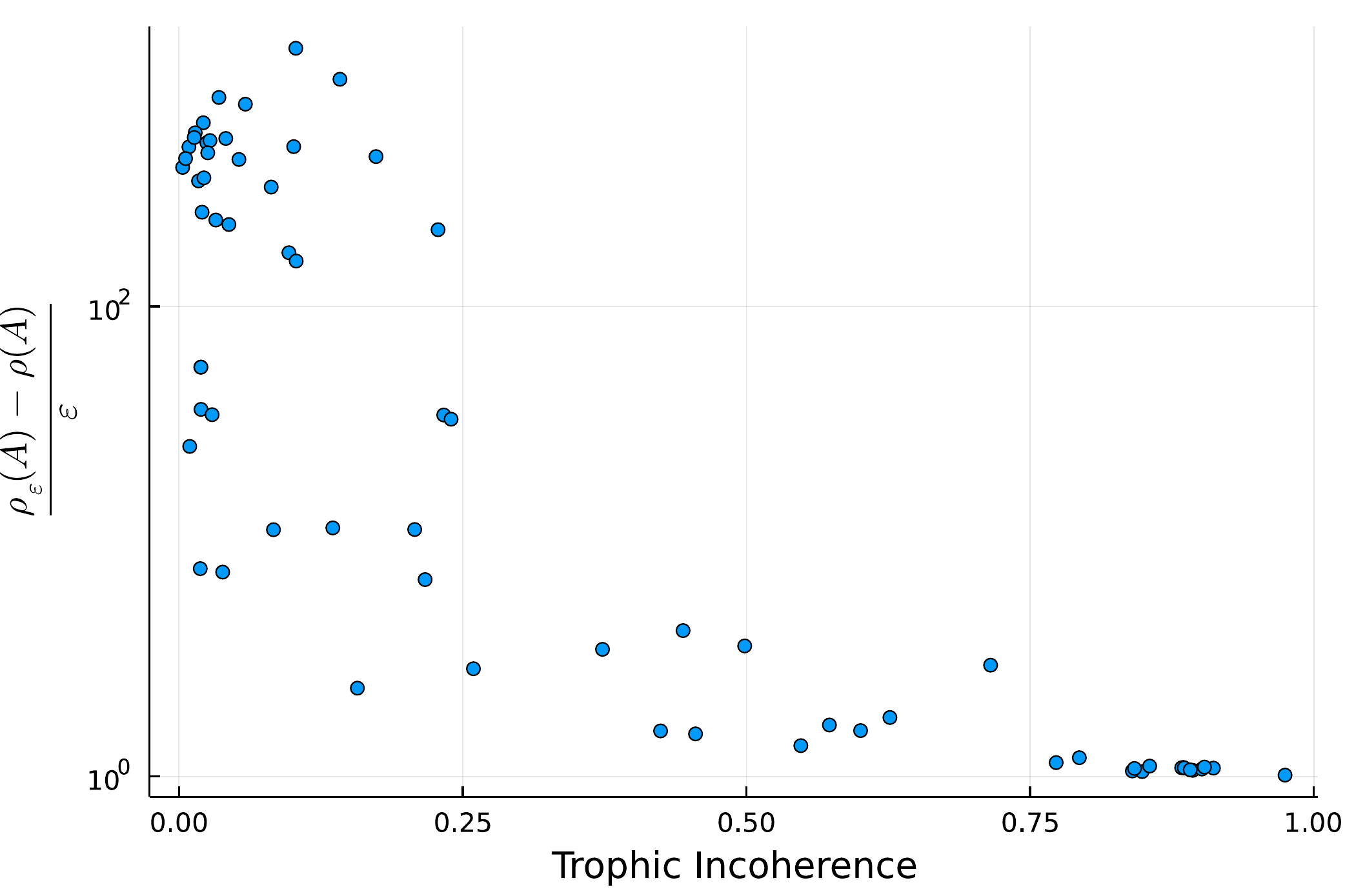}
        	\caption{
        	\textcolor{black}{Real Networks- Taken from \cite{DataSamJohnson} and excluding genetic networks for run time. Each point representing a single network. Spectral perturbation size $10^{-3}$}.}
        	
        \label{fig:pseudospectral_real}
        \end{subfigure}     
   \caption{\textcolor{black}{Pseudospectral Radii scaled by perturbation size with Trophic Incoherence for numerically generated networks and real-world networks \cite{DataSamJohnson}}}
        \label{fig:pseudo_exmaples}
\end{figure}

One way to measure the sensitivity to perturbation is to compute the pseudospectral radius, the spectral radius of the perturbed matrix \textcolor{black}{minus} the original spectral radius, and divide by the size of the perturbation. 
If this quantity is if of order $10^0$ then the system is stable to perturbations of that scale.
The size of perturbation used can be very small. In the following figures the perturbation is of order $10^{-3}$. \textcolor{black}{The pseudospetral radius} exhibits a transition as trophic incoherence is varied for generated networks, figure \ref{fig:pseudospectral_radius_100} and \ref{fig:pseudospectral_radius_500}, \textcolor{black}{being large for networks of low incoherence and then decreasing as incoherence increases, as we would expect.}. A similar result can be found for \textcolor{black}{real-world} networks, figure \ref{fig:pseudospectral_real}.

\textcolor{black}{We again produce a similar trend for both the real and generated networks, with the real network data appearing more noisy, as expected.
As with the various dynamics described above, the pseudospectral radius demonstrates an increased sensitivity to perturbations -- in this case structural -- as networks become more coherent.
}

\subsection{Out-Components}

\begin{figure}[H]
      		\centering
            \includegraphics[width=0.8\linewidth]{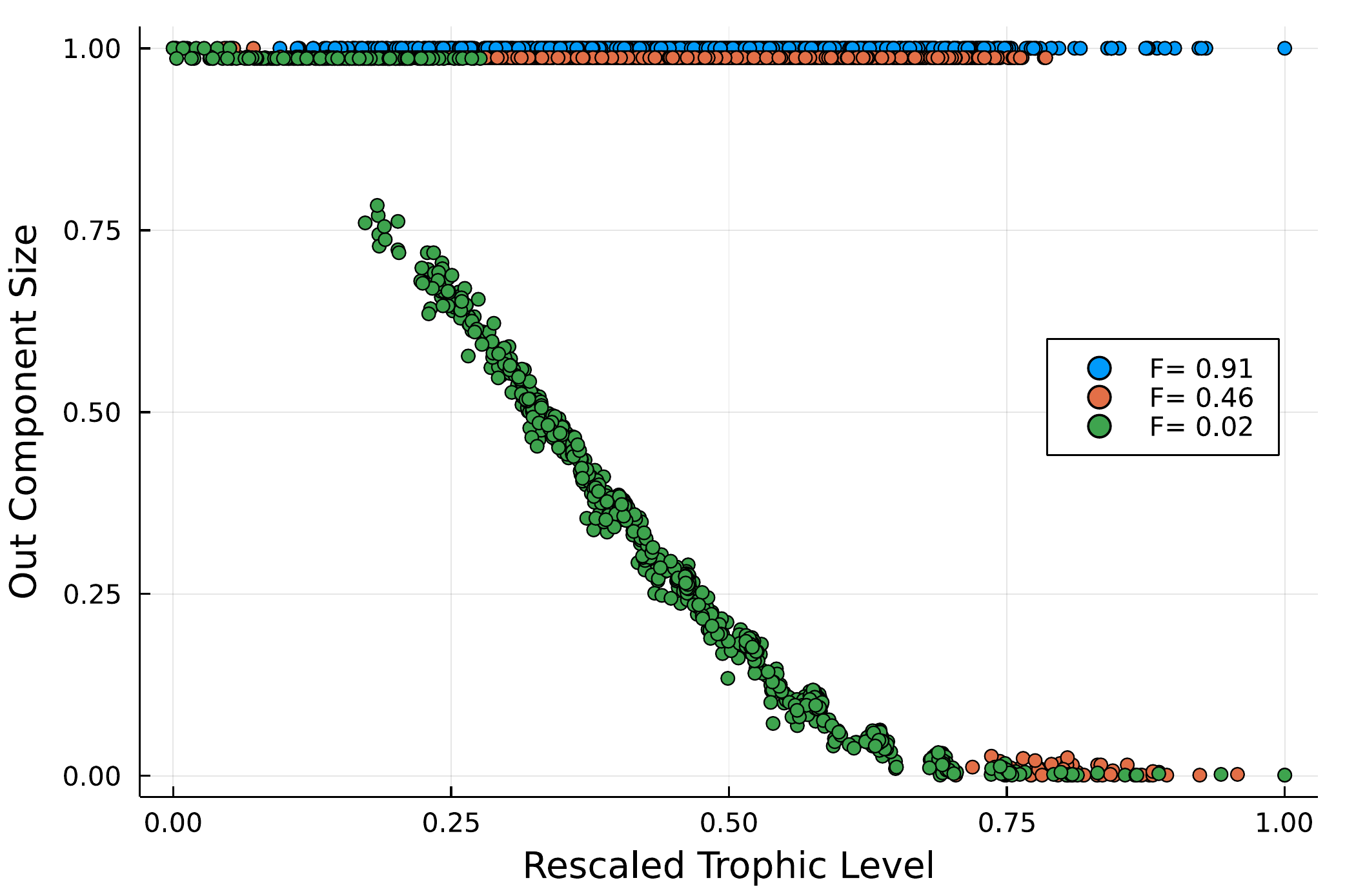}
        	\caption{ Size of the Out-Components of nodes with \textcolor{black}{their} trophic level rescaled by the maximum trophic level. \textcolor{black}{Networks, generated by the model in section \ref{Net_gen} at a single temperature,} of low, intermediate and high incoherence with $N=1000$ and $\langle k \rangle =10$. \textcolor{black}{And $T_{\text{GEN}} = 0.02,1,100$ for} the low, intermediate and high incoherence networks, respectively.
        	}
        	
        \label{fig:out_components}
        \end{figure}  

One way to understand the transition in influence of nodes with hierarchical ordering is to look at the number of nodes which are reachable from a \textcolor{black}{a given node -- its `out component' --} and how that changes with the trophic level \textcolor{black}{of the node}, and \textcolor{black}{with the trophic} coherence of the network. This is shown in figure \ref{fig:out_components}.

When the network is very incoherent \textcolor{black}{it} is strongly connected and there is no heterogeneity in the ability of one node to influence other nodes. When the incoherence is intermediate there is still a large \textcolor{black}{strongly} connected component, however the nodes at the top of the network
\textcolor{black}{are unable to}
influence the network as they lie outside this component. When the network is very coherent and $F$ is close to zero, the strongly connected component is small and the size of \textcolor{black}{the node's} out-component is strongly linked to its 
\textcolor{black}{trophic level.} This leads to an asymmetry in the network due to the hierarchy
\textcolor{black}{of trophic levels,}
where some nodes can reach the whole network whereas others have a very small out-component. These results are \textcolor{black}{a simple representation} since the precise size of the out-component can \textcolor{black}{vary} widely depending on the type of network and how cycles are distributed throughout it. However the general principle that \textcolor{black}{networks} of low incoherence lead to an asymmetry in component size is generally true due to the lack of feedback, 
and agrees with results from the literature \cite{DelaCruzCabrera2019AnalysisExponential}. This is also in agreement with other work \textcolor{black}{on Trophic Analysis and the intuition behind it,} which \textcolor{black}{has analytically linked} the \textcolor{black}{emergence} of  \textcolor{black}{a giant strongly} connected component in a network to its trophic properties \cite{Rodgers2023StrongNetworks}.

\section{Discussion and Conclusion}

These results provide a different perspective on influence \textcolor{black}{and influenceability} in directed complex networks by using the fact that directed networks have an underlying \textcolor{black}{hierarchical structure}, unlike the undirected case.
\textcolor{black}{Thus, Trophic Analysis can give some}
insight into the function of \textcolor{black}{a complex system. It reveals} the links between many network properties, \textcolor{black}{and how a node's trophic level can determine its ability  to influence the rest of network.}

\textcolor{black}{Trophic Analysis
is based on a simple calculation,}
a linear equation dependent only on the degrees and the adjacency matrix, and \textcolor{black}{is straightforward to} interpret. We hope that highlighting the diverse ways it can \textcolor{black}{shed light on} network function and structure \textcolor{black}{might} prompt its use in future work. In particular, when applied to real-world networks for analysis of structural features or understanding of dynamics, as we expect the phenomena we \textcolor{black}{describe} here to be quite general and applicable to many real-world systems. \textcolor{black}{Since the equation is linear it can be solved for large real-world networks.
Additionally, the results can be related to network properties such as }
the spectral radius,
non-normality or
strong connectivity
\cite{MacKay2020HowNetwork,Rodgers2023StrongNetworks}.
\textcolor{black}{
In particular, networks of low trophic incoherence are highly non-normal. It has been shown recently that non-normal networks are very common \cite{Asllani2018StructureNetworks}, that non-normality of the adjacency matrix can have a large effect on the stability of a system \cite{Duan2022NetworkSystems}, and it has been related to economic bubbles \cite{Sornette2022Non-NormalBubbles}. 
}


The general principle that the influence of a node is a function of its position in a hierarchy does not necessarily require the use of trophic levels to measure this and another metric such as PageRank could provide a similar result. However, trophic analysis allows a local measure of node importance to be \textcolor{black}{related to} trophic coherence, which gives a measure of how important the hierarchy is expected to be to the system. \textcolor{black}{This is useful when Trophic Analysis is applied to real-world systems, such as when it has been used to investigate the hierarchy in a network of Sustainable Development Goals \cite{Dawes2022SDGHierarchies,Dawes2022System-levelLevel}.}
This paper highlights how the importance of \textcolor{black}{node-level measures of centrality (such as trophic level or PageRank)}
depends strongly on how these measures are distributed amongst the nodes, as well as on global network properties
\textcolor{black}{(trophic coherence)} which may hinder or accentuate their \textcolor{black}{effects}. Trophic Analysis can be a useful way to understand directed networks as it does not simply rely on either the in or out degree. In a directed network the ability to influence or be influenced may be of different relative importance depending on the dynamics, and
it is possible that no correlation (nor more complex statistical relationship) \textcolor{black}{exists} between in and out degree. So nodes of importance in one measure may not be important in another. Trophic level simplifies this as the low level nodes are good at emitting and have many nodes downstream of them while nodes at the top of the network receive \textcolor{black}{many} paths from below. The interplay between \textcolor{black}{trophic levels} and degree distributions is something that could have a large impact on influence and \textcolor{black}{will} be the subject of future work. For example, a network could have a scale free distribution in either in or out degrees or both, which may or not be correlated, or bear some relationship to the \textcolor{black}{trophic structure}.

In conclusion, we have highlighted how there are many disparate ways to view influence and influenceability \textcolor{black}{in real-world and generated networks,} such as the ability to \textcolor{black}{shape various discrete and continuous} dynamics, the localisation of eigenvectors, the sensitivity of the spectra to \textcolor{black}{structural} perturbations, or the distribution of out-components within networks. We have shown that all these phenomena can be thought of in terms of the local placement of nodes within the hierarchical structure \textcolor{black}{of trophic levels, which is mediated by the global directionality given by trophic coherence}. \textcolor{black}{And we have described how this} insight can be used to understand many structural and dynamical processes in directed complex networks, \textcolor{black}{which we hope prompts the use of Trophic Analysis in the study of specific real-world systems.}

\section*{Acknowledgements}

 The authors would like to thank the Centre for Doctoral Training in Topological Design and Engineering and Physical Sciences Research Council (EPSRC) for funding this research. SJ also acknowledges support from the Alan Turing Institute under EPSRC Grant EP/N510129/1.

\bibliographystyle{unsrt}
\bibliography{references,ref1,extra_ref}

\end{document}